# A Scoping Review of LLM-as-a-Judge in Healthcare and the MedJUDGE Framework

Chenyu Li MS[1,2] ; Zohaib Akhtar MD MPH MBA[3,4] ; Mingu Kwak PhD[2] ; Yuelyu Ji MS [5] ; Hang Zhang MS[5]; Tracey Obi BS[2] ; Yufan Ren BS[2] ;  Xizhi Wu MS[2] ; Sonish Sivarajkumar PhD[5]; Harold P. Lehmann MD PhD[6] ; Shyam Visweswaran MD PhD[1,5,7]; Michael J. Becich MD PhD[1]; Danielle L. Mowery PhD, MS, MS, FAMIA,FACMI[8] ; Renxuan Liu BS[9];  Haoyang Sun[2], Yanshan Wang, PhD[1,2,5,7]*

1. Department of Biomedical Informatics, School of Medicine, University of Pittsburgh, Pittsburgh, PA, USA

2. Department of Health Information Management, School of Health and Rehabilitation Sciences, University of Pittsburgh, Pittsburgh, PA, USA

3. OpenCura, Health Innovation Consortium

4. Northwestern University, Kellogg School of Management

5. Intelligent Systems Program, School of Computing and Information, University of Pittsburgh, Pittsburgh, PA, USA

6. Johns Hopkins University School of Medicine Biomedical Informatics and Data Science, Baltimore, MD, USA

7. Clinical and Translational Science Institute, University of Pittsburgh, Pittsburgh, PA, USA

8. Institute for Biomedical Informatics, University of Pennsylvania, Philadelphia, PA, USA

9. Data Science, School of Computing and Information, University of Pittsburgh, Pittsburgh, PA, USA

*Corresponding author: yanshan.wang@pitt.edu

## Abstract
As large language models (LLMs) increasingly generate and process clinical text, scalable evaluation of their results has become critical. LLM-as-a-Judge (LaaJ), which uses LLMs to evaluate LLM outputs, offers an alternative to costly expert review, yet its healthcare adoption raises safety and bias concerns. We conducted a PRISMA-ScR scoping review of six databases (January 2020–January 2026), screening 11,727 studies and including 49 for this article. The landscape was dominated by evaluation and benchmarking applications (n=37, 75.5%), pointwise scoring paradigms (n=42, 85.7%), and GPT-family judges (n=36, 73.5%). Despite this growing adoption, validation rigor was limited: among the 36 studies with any human involvement, the median number of expert validators was 3; 13 studies (26.5%) used none. Risk of Bias testing was absent in 36 studies (73.5%), only 1 (2.0%) assessed demographic fairness, and none assessed temporal stability or patient context incorporation. Deployment remained nascent, with 1 study (2.0%) achieving production deployment and four (8.2%) reaching prototype stage. Critically, these gaps interact: when judges and evaluated systems share training data and model architecture, they inherit the same knowledge gaps, and standard agreement metrics cannot distinguish accurate evaluation from shared ignorance. Together, minimal human oversight, absent bias testing, and model monoculture constitute a systemic governance failure in which current validation practices cannot detect errors shared between judge and evaluated system — the failures most likely to cause patient harm. To address this, we propose MedJUDGE (Medical Judge Utility, De-biasing, Governance and Evaluation), a risk-stratified three-pillar conceptual framework organized around validity, safety, and accountability across three clinical risk tiers — providing the first deployment-stage evaluation guidance for healthcare LaaJ systems.

## Introduction

Large language models (LLMs), representing the state-of-the-art in natural language processing (NLP), are rapidly transforming healthcare by supporting clinical documentation[1], triage[2], question answering[3], and increasingly, enabling autonomous medical agents[4] capable of orchestrating multiple tools. Because LLMs may directly influence clinical decision-making and patient outcomes, rigorous evaluation of LLMs in healthcare settings has become essential.

The NLP and computational linguistics communities have historically developed extensive quantitative metrics, such as Bilingual Evaluation Understudy (BLEU)[5], Recall-Oriented Understudy for Gisting Evaluation (ROUGE)[6], and perplexity scores, for assessing traditional NLP models. However, as LLMs generate novel and complex content, these traditional metrics have proven insufficient for assessing critical dimensions of healthcare performance, including clinical readiness, reasoning quality, and patient safety implications[7,8].

Consequently, human expert evaluation has emerged as the de facto gold standard for validating LLMs in clinical contexts. Yet, this reliance on manual evaluation introduces a central bottleneck in healthcare.[9,10] It is resource-intensive, time-consuming, and fails to scale with the rapid influx of LLM-generated content in routine clinical workflows[11,12]. The need for scalable evaluation is further amplified by emerging global regulatory demands. Specifically, the FDA's Predetermined Change Control Plan (PCCP) final guidance[13] mandates ongoing evaluation for adaptive AI including LLMs; the EU AI Act[14] classifies medical AI and LLMs as "high-risk" to enforce stringent transparency and human oversight; and the WHO explicitly warns against hallucination risks and automation bias[15]. Together, these frameworks necessitate continuous, scalable monitoring systems.

To address this evaluation bottleneck, researchers at the intersection of clinical NLP and AI evaluation have begun adopting LLM-as-a-Judge (LaaJ), using high-performing models to evaluate the outputs of others[16]. In general domains, LaaJ has enabled automated benchmarking with 70–90% human agreement. However, LaaJ introduces systematic evaluation risks, including positional bias[17], verbosity bias[18], self-preference bias[19], calibration errors [20], and susceptibility to adversarial manipulation[21,22]. While well-documented in general text generation, these risks remain largely unexamined in clinical contexts. The field lacks a comprehensive review of LaaJ approaches used to evaluate LLMs in healthcare applications, creating a significant gap in guidance for safe evaluation architectures in medicine. We argue that healthcare LaaJ systems face a governance failure invisible to existing frameworks: evaluation errors compound through multi-judge pipelines, amplify demographic disparities through shared training blind spots, and propagate unchecked through hidden evaluation components embedded throughout clinical AI pipelines, problems that cannot be solved by improving individual judge quality alone.

Several recent reviews focus on LLM evaluation in healthcare, but none on the LaaJ frameworks in clinical contexts. Lee et al [23] conducted a scoping review of evaluation methods for medical LLMs, analyzing 142 full-text articles across clinical specialties. Their review cataloged evaluation dimensions, sample sizes, and statistical approaches but focused exclusively on human evaluation methodologies, how physicians and other experts assess LLM outputs, and did not examine automated evaluation by LLM judges. Similarly, Shool et al. [24] systematically reviewed 761 studies evaluating LLMs in clinical medicine, characterizing performance benchmarks, clinical tasks, and outcome metrics. Their analysis concentrated on what was being evaluated (LLM clinical performance) rather than who or what was performing the evaluation, and their corpus contained no instances of LLM-as-a-Judge. Tam et al.[25] reviewed 142 studies to develop the QUEST framework for human evaluation of healthcare LLMs, proposing standardized dimensions (Quality of Information, Understanding and Reasoning, Expression Style and Persona, Safety and Harm, and Trust and Confidence) and practical guidance for evaluator recruitment and statistical analysis. While QUEST provides essential infrastructure for human evaluation, it explicitly acknowledges the limitations of relying solely on human reviewers at scale and does not address the emerging practice of substituting or augmenting human judges with LLM-based evaluators. In the non-medical domain, Gu et al.[16] and Li et al.[26] have produced comprehensive surveys of LaaJ methodology, documenting evaluation paradigms, known biases — including positional bias (scores shifting based on answer order), verbosity bias (longer responses rated higher regardless of quality), and self-preference bias (a model favoring outputs from its own family) — and mitigation strategies. But they do not account for the unique safety-critical implications of evaluation errors in healthcare. More fundamentally, it remains unknown whether LLM judges in healthcare have access to or appropriately incorporate the patient-level context, comorbidities, medication history, and prior interventions that determine whether a clinical recommendation is safe for a specific patient. Without such access, automated evaluation may systematically miss the most dangerous errors.

To bridge this gap, we conducted a scoping review that systematically analyzes the emerging landscape of LaaJ applications in healthcare. We synthesize methodologies, evaluation paradigms, validation rigor, bias mitigation practices, and deployment readiness across published studies. Specifically, the contributions of this review are the following:

(1) systematically analyzing LaaJ methodologies and applications in healthcare, characterizing evaluation paradigms, model architectures, validation practices, and deployment patterns;

(2) studying LaaJ validation rigor, bias assessment, robustness testing, and deployment readiness; and

(3) synthesizing findings into MedJUDGE, a risk-stratified evaluation framework providing the first deployment-stage evaluation guidance for healthcare LaaJ systems.

## Results

### Study selection results

Searches across six electronic databases (Ovid MEDLINE and Epub Ahead of Print, Ovid EMBASE, Scopus, Web of Science, ACM Digital Library, and IEEE Xplore) yielded 12,379 citations. After removing 652 duplicates, 11,727 unique citations underwent screening. Two independent reviewers (TO and YR) screened titles and abstracts, with full-text assessment against inclusion/exclusion criteria identifying 33 eligible articles. Supplementary citation screening of included articles identified 16 additional eligible studies, for a final corpus of 49 included articles. The complete selection process is documented in the PRISMA-ScR flow diagram[27] ( see Fig. 1).

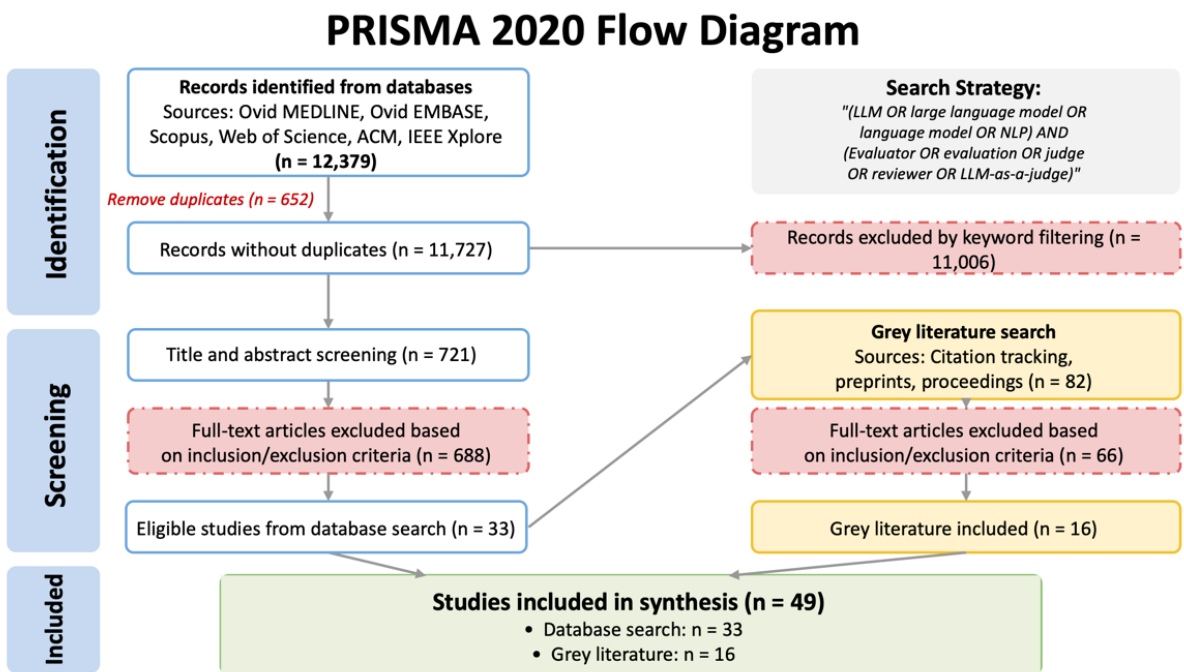

*Figure 1 PRISMA-ScR flow Diagram*

The 49 included studies span 2023–2026, with pronounced temporal clustering: 1 study in 2023 (n = 1, 2.0%), 12 in 2024 (n = 12, 24.5%), and 36 in 2025–2026 (n = 36, 73.5%), reflecting the rapid recent emergence of healthcare LaaJ applications (Supplementary Fig. 1). The grey literature component (n= 16, 33%) reflects the field's early stage, with much work circulating as preprints or recent conference presentations not yet indexed in major electronic databases.

We analyzed the 49 included articles across five dimensions: (1) application categories reflecting how LaaJ is deployed in healthcare workflows, (2) clinical tasks and specialty distributions, (3) evaluation paradigm selection and its relationship to task complexity, (4) judge model selection and adaptation strategies, and (5) validation rigor, including human expert involvement, bias testing, and deployment readiness. We first present the application landscape, then examine methodological patterns, and conclude with validation gaps that motivate the MedJUDGE framework.

**Application Categories of LaaJ**

Based on our synthesis of 49 articles, we identified three distinct application categories (Supplementary Tables 1–3) reflecting how LaaJ is used in healthcare; counts, definitions, and representative studies are summarized in Supplementary Table 4 and Supplementary Fig. 2.

***Category 1: Evaluation and Benchmarking (n = 37, 75.5%).*** In this category, LLM judges serve as standalone assessment tools that score or compare outputs without influencing model training. Performance variability reflected task complexity: structured documentation tasks achieved the highest agreement (0.82–0.93)[28–31], benchmark tasks showed moderate agreement (0.64–0.79)[32–35] and diagnostic reasoning tasks varied most widely (0.47–0.90), with agreement constrained by human expert disagreement.

***Category 2: Reinforcement Learning (n = 3, 6.1%).*** In this category, LLM judges serve as reward models whose scores directly guide model training; here, the judge's ratings become training signals that shape how the evaluated model improves. Two training approaches appear: Reinforcement Learning from Human Feedback (RLHF)[36], where models are fine-tuned using human preference judgments, and Reinforcement Learning from AI Feedback (RLAIF)[37], where LLM-generated preferences replace human annotations. A related technique, Direct Preference Optimization (DPO)[38], achieves similar alignment through a simplified training procedure that eliminates the explicit reward model.

***Category 3: A Component in a Multi-Agent System (n = 9, 18.4%).*** The third category of LaaJ is that LLM judges are being used as a component in a multi-agent system to evaluate outputs of agents and provide feedback to improve agents. The LLM judges are being used in several forms: 1) heterogeneous judge panels, where models from different model families (such as the GPT family from OpenAI or the Claude family from Anthropic) independently evaluate the same output and aggregate their scores (analogous to a multidisciplinary tumor board reaching consensus from diverse specialist perspectives); 2) hierarchical agent structures[39], where specialized agents handle distinct evaluation subtasks (e.g., one agent verifies factual accuracy, another assesses safety, a third evaluates communication quality) before an orchestrator integrates their assessments; 3) meta-evaluation[26], where a judge evaluates the quality of another judge's assessment, adding a layer of quality control to the evaluation process itself; and 4) ensemble approaches, where multiple instances of the same or similar models evaluate independently and their outputs are combined through voting or averaging to reduce individual model variance.

The concentration of 75.5% of studies in evaluation-only applications, with no study progressing beyond prototype in training or autonomous decision-making pipelines, reflects not technical immaturity, but the absence of any framework specifying what safe progression would require.

**Clinical Tasks and Specialty Distributions**

Studies are clustered into three primary application domains: clinical content evaluation, medical knowledge and reasoning assessment, and patient-facing and decision support applications (details see Supplement Table 12). **Clinical content evaluation** (n = 17, 34.7%) encompassed clinical documentation and summarization (n = 7, 14.3%), report generation (n= 4, 8.2%), and information extraction (n= 6, 12.2% ). **Medical knowledge and reasoning assessment** (n = 17, 34.7%) included medical QA and reasoning (11 studies) and diagnostic support (6 studies). **Patient-facing and decision support applications** (n = 12, 24.5%) covered patient education and communication (6 studies) and clinical decision support systems (6 studies). Three studies did not fit neatly into these categories.

Nearly half the studies (n=22, 44.9%) addressed multiple specialties or general medicine. Among single-specialty studies, radiology was most common (n= 6, 12.2%), followed by endocrinology/diabetology (n= 4, 8.2%), cardiology (n=3, 6.1%), and plastic/hand surgery (n=3, 6.1%). Notable gaps include emergency medicine, pediatrics, and surgery, fields where clinical decision-making is time-sensitive and high-stakes.

**Evaluation Paradigm Selection and its Relationship to Task Complexity**

**Table 1. Key evaluation paradigms in LaaJ**

| Paradigm | Definition | Computational cost | Best suited for |
|---|---|---|---|
| Pointwise scoring[40] | Assigns absolute scores or categorical labels to a single output using predefined criteria, analogous to grading an essay against a rubric | $O(n)$ | Structured tasks with established quality frameworks |
| Pairwise comparison[41] | Presents two candidate outputs side-by-side to determine which is superior | $O(n^2)$ | Subjective quality comparison with few candidates |
| Agent-based evaluation[42] | Decomposes assessment into modular subtasks handled by specialized components | 3–10× pointwise | Complex, safety-critical tasks requiring decomposition |

Table 1 lists three evaluation paradigms used in the reviewed articles as well as their definitions, computational costs, and best-suited applications. (Supplementary Tables 5) Pointwise or rubric-based scoring dominated the landscape: 42 of 49 studies (85.7%) employed this paradigm as their primary evaluation method (Table 1; Supplementary Fig. 3). Pairwise comparison appeared in 7 studies (14.3%), while agent-based multi-step evaluation characterized 6 studies (12.2%). A clear inverse relationship emerged between task structure and agreement: structured documentation tasks achieved 0.82–0.93 agreement, semi-structured assessments reached 0.64–0.79, and complex reasoning tasks showed 0.47–0.71. Notably, for complex reasoning tasks, model–physician agreement barely differed from physician–physician agreement (HealthBench: model–physician MF1 = 0.572–0.706 vs. physician–physician MF1 = 0.569–0.730; MedHELM: LLM-jury ICC = 0.47 vs. clinician–clinician ICC = 0.43)[43,44], establishing fundamental performance ceilings constrained by legitimate medical expert disagreement.

**Judge Model Selection and Adaptation Strategies**

There are eight LLMs that are used as judges: GPT-family models, Claude, Llama, Qwen, Gemini, Deep Seek, Mistral, and BERT models. GPT-family models dominated as judges, used in 36 of 49 studies (73.5%), with 27 using GPT as the sole judge family (Supplementary Fig. 4, Supplementary Tables 8). Claude models served as judges in 6 studies (12.2%), primarily as cross-validation judges in multi-model designs. Llama models appeared in 8 studies (16.3%), and Gemini in 4 (8.2%). The majority of judges were general-purpose: 43 of 49 studies (87.8%) used judges without medical specialization. Only 6 studies (12.2%) employed judges with domain adaptation, including supervised fine-tuning[45], medically-specialized discriminative models[46], expert persona prompting[47], and domain-specialized prompting for radiology[33] (Supplementary Table 6).

Adaptation methods included prompt engineering (n=15 studies, 30.6%), few-shot/in-context learning (n=12, 24.5%), supervised fine-tuning (n=11, 22.4%), retrieval-augmented generation (n=8, 16.3%), chain-of-thought (n=4, 8.2%), LoRA/QLoRA (n=4, 8.2%), and DPO/RLAIF (n=3, 6.1%) (Supplementary Tables 7).

Among studies employing structured prompt engineering, four (8.2%) implemented G-EVAL[48], the general-domain evaluation architecture that pairs rubric-based GPT-4 scoring with auto-generated chain-of-thought reasoning steps to improve human alignment. All four applied G-EVAL within pointwise scoring designs; three[49–51] (Ozmen et al. 2025a, 2025b, 2025c) used it to evaluate RAG-based surgical clinical decision support systems, and one[52] applied it to radiology report classification with KNN-based dynamic few-shot retrieval. Only 8.2% of studies adopted this framework despite documented reliability gains in general-domain evaluations (Spearman $\rho$ up to 0.51), reflecting the broader gap between general-domain NLP methodology and clinical LaaJ practice.

Using the same model family as both generator and judge creates a specific problem: both systems share the same training data and therefore the same knowledge gaps. A GPT-based judge evaluating GPT-generated clinical text will score highly on errors that GPT-family models systematically miss, not because those errors are absent, but because neither system recognizes them as errors.

**Validation Rigors**

Because LaaJ systems are themselves LLM models, their validity cannot be assumed. We characterize validation practices across two dimensions: bias testing and mitigation, and human expert involvement.

*Bias testing and mitigation:* Bias testing remains the most significant methodological gap. Only 13 studies (26.5%) conducted any form of bias testing, while 36 (73.5%) reported none (Table 2; Supplementary Fig. 5, Supplementary Table 10). Among the 13 studies with testing, only 8 (16.3%) tested specific named bias dimensions. Positional bias was tested in 3 studies (6.1%), self-enhancement in 2 (4.1%), and length/verbosity in 1 (2.0%). Only one study[44] assessed demographic bias across race, gender, or socioeconomic status. Only 3 studies (6.1%) formally assessed equity dimensions.

**Table 2. Bias dimensions tested across included studies (N = 49)**

| Bias dimension | Studies testing, n (%) | Clinical risk if untested |
| --- | --- | --- |
| Positional | 3 (6.1%) | High |
| Length/verbosity | 1 (2.0%) | Medium |
| Format | 0 (0%) | Medium |
| Self-enhancement | 2 (4.1%) | Critical |
| Demographic | 1 (2%) | Critical |
| Severity calibration | 0 (0%) | High |
| Temporal stability | 0 (0%) | High |
| Cross-specialty | 2 (4.1%) | High |

*Human expert involvement:* Human expert involvement varied widely (Supplementary Fig. 6). The median number of human evaluators was 2 (IQR: 0–5) across all studies and 3 (IQR: 2–7) among the 36 studies with human involvement. We stratified expert involvement into four levels: extensive (≥20 evaluators, 5 studies, 10.2%), moderate (5–19, 10 studies, 20.4%), minimal (1–4, 21 studies, 42.9%), and none (13 studies, 26.5%) (Details see supplementary Table 9). Among studies with extensive panels, HealthBench[43] engaged 262 physicians and MedHELM[44] recruited 49 clinicians spanning 14 specialties across 4 institutions. Expert involvement took three forms: reference standard creation, where experts generated or annotated gold-standard outputs against which judge scores were compared[23,25]; direct rating, where experts scored the same outputs as the judge using Likert scales or binary accept/reject judgments to compute agreement[9,25]; and post-hoc review, where experts assessed a sample of judge outputs to validate overall quality without item-level comparison[10]. Reference standard creation represents the strongest form of validation, as it provides an independent anchor for judge performance; direct rating and post-hoc review are weaker because they conflate judge reliability with task difficulty. Expert credentials ranged from board-certified specialists with 10–30 years of experience to graduate students in biomedical fields; 6 studies (12.2%) did not specify validator credentials.

Without bias testing, there is no way to know whether a judge's consistent scores reflect genuine quality or a consistent, undetected error. Consistency and accuracy are not the same thing.

## The MedJUDGE Framework: A Risk-Stratified Framework for Healthcare LaaJ

In this review study, we have identified several gaps in using LaaJ for healthcare applications, including, but not limited to, inadequate validation, absent bias testing, unquantified error propagation, and model monoculture. Existing frameworks are usually not sufficient to address these gaps. For example, MedHELM provides pre-deployment benchmarking,[44] and QUEST provides human evaluation design[25], but neither covers continuous validation during live clinical deployment, nor accounts for bias testing, or quantifying error propagation in the frameworks. To address these gaps, we propose MedJUDGE (Medical Judge Utility, De-biasing, Governance and Evaluation): a risk-stratified evaluation framework for applying LaaJ in healthcare applications. Here, Governance refers to the operational accountability mechanisms, including expert validation requirements, post-deployment monitoring, and audit trail documentation, through which institutions establish and maintain the trustworthiness of LaaJ systems throughout their clinical lifecycle. MedJUDGE thus represents empirically grounded propositions derived from observed practical gaps in this review study.

Unlike a sequential pipeline, MedJUDGE operates as a risk-tier-gated requirements matrix. First, users of MedJUDGE need to classify their LaaJ deployment by clinical risk tiers. Second, according to the clinical risk tier, users need to address requirements across three pillars, including (1) Validity, (2) Safety, and (3) Accountability (see an overview in Figure 2). In the following sections, we describe the clinical risk tiers and the three foundational pillars that guide the use of MedJUDGE.

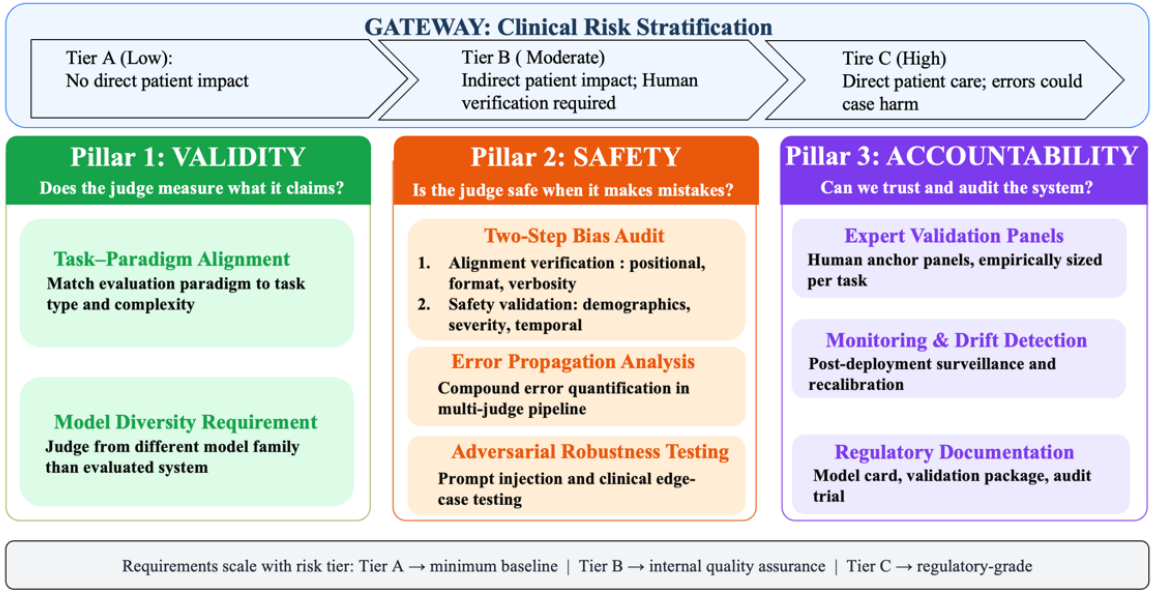

Figure 2 MedJUDGE Framework

**Clinical Risk Tiers**

Healthcare applications adopting MedJUDGE are classified into one of three clinical risk tiers as an entry point: all subsequent framework requirements are determined by the tier assigned at this stage. Table 3 summarizes the definitions, examples, and oversight analogues of each tier.

**Tier A (Low Clinical Risk) applications** correspond to research and educational use cases in which LaaJ outputs do not directly influence patient care decisions. Errors at this tier primarily affect knowledge generation, research interpretation, or educational outcomes rather than clinical actions. Typical examples include research question answering, automated literature summarization, hypothesis generation, and educational training support. Validation requirements at this tier emphasize transparency of model limitations, appropriate disclosure of AI assistance, and adherence to institutional data use policies.

**Tier B (Moderate Clinical Risk) applications** include clinical support activities that may indirectly influence patient care and require mandatory human verification before any clinical use. These systems function as clinical workflow augmentation tools rather than autonomous decision-makers. Examples include clinical documentation quality assurance, EHR note summarization for clinician review, structured extraction of laboratory results and medication lists to support reconciliation workflows, risk-based patient pre-screening, and structured data extraction to support quality measurement. Errors at this tier could affect clinical efficiency or documentation quality: human verification therefore serves as the primary safety control for mitigating downstream patient risk. Validation requirements at this level include institutional oversight, performance monitoring for drift, and clearly defined human-in-the-loop procedures.

**Tier C (High Clinical Risk) applications** involve use cases in which LaaJ outputs may directly inform or influence patient care decisions and therefore carry the highest level of clinical risk. These applications include clinical decision support for triage, diagnostic reasoning assistance, and medication safety alerts, where errors could result in patient harm if not properly mitigated. Because of their potential impact on clinical decision-making, these systems may fall within the scope of Software as a Medical Device (SaMD) and may require regulatory oversight consistent with FDA guidance, depending on their intended use and level of clinical autonomy. Validation requirements at this tier include rigorous pre-deployment evaluation, prospective assessment in real-world clinical settings, and formal risk management processes. Institutions adopting Tier C applications should establish continuous post-deployment monitoring mechanisms and implement safeguards such as clinician override capabilities and audit trails to ensure safe deployment.

Notably, all 49 studies included in our review remained within low-to-moderate risk categories, with only 2.0% achieving production deployment and 8.2% reaching the prototype or proof-of-concept stage. The clinical risk classification of an application may also evolve over time: institutions are encouraged to adopt a staged implementation strategy, beginning with Tier A applications and progressing to higher-risk tiers only after satisfying the validation and monitoring requirements defined by MedJUDGE at each level. As clinical risk increases, the corresponding MedJUDGE evaluation requirements should be applied at the appropriate rigor level to ensure continued safe and compliant use.

**Table 3. Clinical risk tiers in MedJUDGE: definitions, examples, and oversight analogue.**

| Clinical Risk Tier | Definition | Examples | Oversight Analogue |
|---|---|---|---|
| Tier A (Low) | No direct patient impact; errors affect research only | Research QA, educational training support, literature summarization | IRB-exempt quality improvement |
| Tier B (Moderate) | Indirect patient impact; mandatory human verification | Clinical documentation QA, EHR summary assessment, patient pre-screening, labs extraction | Institutional quality assurance / peer review |
| Tier C (High) | Direct patient care decisions; errors could cause harm | Triage support, diagnostic reasoning evaluation, | FDA SaMD oversight |

| Clinical Risk Tier | Definition | Examples | Oversight Analogue |
|---|---|---|---|
| | | medication safety alerts where judge output directly informs prescribing or discontinuation decisions | |

**Requirements Matrix**

Based on the clinical risk tier, MedJUDGE organizes evaluation requirements into three complementary pillars: (1) **Validity**, (2) **Safety**, and (3) **Accountability**. Figure 3 provides an overview of the complete requirements. Together, these pillars define a risk-proportionate evaluation framework to ensure that LaaJ is applied with methodological rigor, clinical safety, and operational accountability throughout its lifecycle. The **Validity pillar** focuses on the methodological rigor of using LaaJ, including task–paradigm alignment and model diversity to ensure a robust and generalizable evaluation approach. The **Safety pillar** addresses risk mitigation through bias assessment, error propagation analysis, and adversarial robustness testing proportional to the level of clinical risk. The **Accountability pillar** ensures operational responsibility through expert validation, performance monitoring, and appropriate regulatory documentation. As clinical risk increases from Tier A to Tier C, the expectations within each pillar become progressively more stringent, reflecting the need for stronger evidence, oversight, and safeguards for LaaJ systems with greater potential impact on patient care. The details for each pillar are described below.

## MedJUDGE Requirements Matrix by Risk Tier

| Requirement | Tier A (Low) | Tier B (Moderate) | Tier C (High) |
|---|---|---|---|
| **PILLAR 1: VALIDITY** | | | |
| Task–paradigm alignment | Pointwise for structured tasks; Pairwise for comparative tasks; Multi-agent for safety-critical tasks | | |
| Model diversity | Different model family preferred | Different model family test | Test on multi-family panel |
| **PILLAR 2: SAFETY** | | | |
| Bias audit    Tier 1 | All 3 tests ; mast pass | | |
| Tier2 | Conduct where feasible | Critical dimensions = blocker | All dimensions + external audit |
| Error propagation | Document if multi-judge | Quantify (Equations 1–4) | Quantify worst-case + error budget |
| Adversarial robustness | Not required | Prompt injection | Full adversarial test |
| **PILLAR 3: ACCOUNTABILITY** | | | |
| Expert validation | Single-site panel | Multi-site specialists | Multi-institution; replication |
| Monitoring & drift | Lock versions; log dates | Quarterly recalibration | Continuous SPC + sentinels |
| Regulatory docs | Model card | Validation package | FDA engagement; audit trail |

Legend: Optional/Document only | Recommended | Required | Required+stricter conditions | Required+external/regulatory

Figure 3 Three Pillar Requirement Matrix by Clinical Risk Tier in MedJUDGE

**Pillar 1: Validity, Does the LaaJ Assess What it Claims to Assess?**

*Task–paradigm alignment.* Paradigm selection must be driven by task characteristics, not convenience. Figure 4 depicts the decision tree of selecting the appropriate paradigm for a given clinical evaluation task.

1. **Determine whether the evaluation task is a structured task.** Structured evaluation tasks, such as guideline adherence checking, documentation completeness assessment, and named entity extraction, are best aligned with pointwise scoring approaches. These tasks typically rely on binary criteria or rubric-based evaluation standards that minimize subjective interpretation and allow efficient scaling across large volumes of outputs. Consistent with this alignment, pointwise scoring demonstrates high inter-rater agreement (0.82–0.93) with linear computational cost ($O(n)$), making it particularly suitable for high-throughput and well-defined clinical evaluation scenarios.
2. **Determine whether the evaluation task is a comparative task.** Comparative tasks that require relative quality assessment rather than absolute scoring, such as evaluating communication quality or explanatory clarity, are better suited for pairwise comparison paradigms. Although pairwise approaches incur higher computational cost ($O(n^2)$), they provide greater reliability for subjective evaluations. Psychometric research in our literature review demonstrates that both humans and LLM judges produce more stable relative judgments than absolute ratings in subjective domains. This advantage makes pairwise comparison particularly valuable for evaluations involving nuanced qualitative differences that may not be well captured by fixed scoring rubrics.
   3. **Determine whether the evaluation task is a complex and safety-critical task.** For complex and safety-critical evaluations, such as diagnostic reasoning assessment, medication safety review, and multi-step triage evaluation, multi-agent decomposition approaches are recommended. These paradigms distribute evaluation into modular subtasks performed by specialized agents, improving transparency and robustness in complex clinical reasoning assessments. Such approaches also require explicit error propagation analysis prior to deployment to ensure that upstream evaluation errors do not lead to unsafe downstream conclusions. Evidence supporting this approach includes MAGI[53], which achieved the highest inter-rater reliability (ICC = 0.87) among studies in our review, demonstrating the value of structured multi-agent evaluation for complex clinical tasks.

Notably, despite these paradigm–task alignment considerations, 85.7% of studies in our review relied on pointwise scoring regardless of task characteristics, which limit evaluation quality and likely contributes to the greater variability in agreement observed for complex reasoning tasks (0.47–0.71), as shown in Supplementary Tables 1 and 5 Following the **Validity pillar** of the MedJUDGE framework, healthcare applications will have improved reliability, interpretability, and reproducibility of LaaJ.

**Paradigm Selection Decision Tree for Healthcare LaaJ**

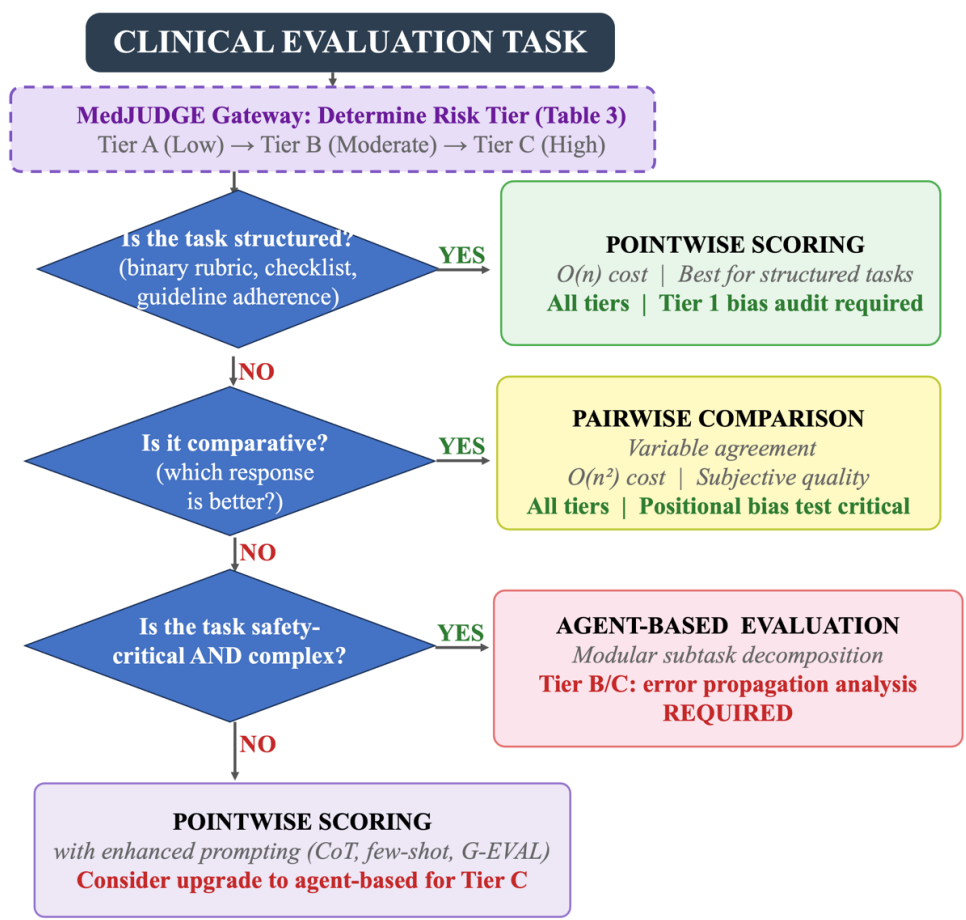

Figure 4. Decision tree of paradigm selection for healthcare LaaJ.

*LaaJ model adaptation.* Once the evaluation paradigm is selected, the adaptation strategy needs to be driven by task requirements and risk tier, not availability. We identified seven model adaptation strategies across the reviewed studies (Supplementary Table 7) and propose the following recommendations for adapting LLM-as-a-Judge (LaaJ) models within the MedJUDGE framework. First, prompt engineering and few-shot learning should be considered minimum requirements across all risk tiers, as they form the foundation for task alignment and evaluation consistency. Notably, 40% of studies did not report their prompt templates, representing a major reproducibility gap and highlighting prompt specification as a primary methodological vulnerability in the current literature. Auto-prompt optimization may also be considered to improve accuracy. Second, we recommend the use of persona assignment when LLMs are used as judges. Depending on the evaluation context, system prompts may incorporate personas such as clinicians, patients, or individuals representing diverse demographic backgrounds (e.g., race or gender) to better reflect real-world clinical perspectives. Our review suggests that incorporating diverse personas can improve the robustness, reliability, and generalizability of evaluation results by reducing perspective bias. Third, retrieval-augmented generation (RAG) represents an important enhancement for judge models, particularly for Tier B and Tier C applications. By dynamically incorporating up-to-date clinical guidelines and institutional policies at inference time, RAG can improve temporal validity and reduce reliance on potentially outdated parametric knowledge embedded in pretrained models. This approach is especially important for clinical domains where standards of care evolve rapidly. Fourth, post-training approaches—including supervised fine-tuning (SFT) (n=11, 22.4%), parameter-efficient methods such as LoRA/QLoRA (n=4, 8.2%), and preference optimization techniques such as DPO or RLAIF (n=3, 6.1%)—can further improve task alignment and evaluation performance. However, these adaptations introduce additional validation requirements: any fine-tuned judge model should undergo independent validation on held-out clinical datasets, and its bias and performance characteristics should be reassessed following each retraining cycle to ensure that model adaptation does not inadvertently introduce performance degradation or new sources of bias.

*Model diversity.* We recommend that judge LLMs be selected from a different model family than the system being evaluated to reduce the risk of self-enhancement bias.[19] Here, model family refers to models that share a common training data origin and alignment pipeline rather than similar inference behavior. For example, GPT-4, GPT-4o, o3, and GPT-5 all belong to the GPT family; LLaMA 2 and LLaMA 3 belong to the LLaMA family; DeepSeek R1 and DeepSeek V3 belong to the DeepSeek family; and Qwen 2 and Qwen 3 belong to the Qwen family, regardless of whether they use extended chain-of-thought reasoning during inference. This distinction is important because models developed within the same family may inherit similar blind spots, stylistic preferences, and alignment tendencies, thereby increasing the risk that shared biases will inflate evaluation agreement or favor outputs generated by related systems. To achieve meaningful model diversity, the judge should therefore come from a different organizational and developmental lineage than the evaluated model. For example, a GPT-family generator should ideally be evaluated by a judge drawn from a different family, such as LLaMA, DeepSeek, or Qwen. For Tier B and Tier C applications, we further recommend the use of multi-family judge panels to reduce the risk of correlated errors across judges. Panel composition should be determined empirically: institutions may begin with a two-family panel, assess inter-judge error correlation on held-out clinical cases, and incrementally expand the panel until the marginal reduction in correlation falls below a predefined threshold. In addition, to support reproducibility and auditability, studies should report exact model version identifiers, API access dates, temperature settings, and random seeds. The GPT-family monoculture observed in our review, in which 73.5% of studies relied on GPT-based judges, represents a systemic methodological vulnerability that this recommendation is intended to address.

## Pillar 2: Safety, Is the LaaJ Safe When It Makes Mistakes?

Pillar 2 of MedJUDGE focuses on the safety evaluation of LLM judges. We propose three key evaluation components: a two-step bias assessment, error propagation analysis, and optional adversarial robustness testing. Together, these evaluations aim to identify systematic biases, quantify the potential impact of judgment errors on downstream decisions, and assess the resilience of judge models against adversarial inputs. These safety evaluations become increasingly stringent with higher clinical risk tiers to ensure that LLM-based evaluation systems do not introduce unacceptable risks into clinical or research workflows.

*Two-step bias testing.* Bias testing in LaaJ includes two steps, alignment verification and safety validation test.

**Step 1: Alignment verification.** A judge that changes its scores because answer order was swapped, or because the scoring scale was reformatted, has not adequately internalized the evaluation task. All downstream assessments from such a judge are unreliable regardless of other properties. Three alignment tests are required at every risk tier (Table 4).

Table 4. Step 1 bias tests of alignment verification

| Test | Protocol | Pass criterion |
|---|---|---|
| Positional bias | Present same content in different orderings | <5% score variance |
| Format bias | Same content evaluated on different scoring scales; compare normalized scores | <5% variance after normalization |
| Length/verbosity bias | Correlate scores with response length on quality-controlled samples | r < 0.2 |

Because no clinical-task-specific calibration data exist, the pass criteria below are proposed as empirically motivated starting points rather than validated standards. For positional and format bias, a pass criterion of less than 5% score variance is adopted, informed by two observations: well-calibrated judges demonstrate near-invariance to ordering under position-swapping conditions[18], and 5% tolerance is consistent with risk-stratified acceptable measurement error in clinical quality assurance, where error tolerance scales with clinical consequences rather than as a universal absolute[54–56]. For length bias, r < 0.2 is adopted (the midpoint of Cohen's small-effect band[57]) at which response length explains less than 4% of score variance ($r^2 < 0.04$), a threshold used in general-domain LaaJ validation as evidence that scoring reflects content quality rather than superficial length features[58,59]. Institutions should treat a borderline pass (e.g., 4–6% variance) as a flag for additional investigation rather than automatic approval.

Empirical calibration across diverse clinical tasks and rubric designs is required before these thresholds can be treated as validated standards.

If any Step 1 test fails, the judge is fundamentally unreliable for the target task and deployment should not proceed. Larger, more recent models exhibit greater alignment stability consistent with general capability scaling trends[60], but alignment must be verified for each specific clinical task and rubric: general-domain consistency does not guarantee domain-specific reliability.

**Step 2: Safety validation test.** Alignment verification does not guarantee safety. A judge may score consistently while exhibiting systematic demographic bias, miscalibration across clinical severity levels, or preferential scoring of outputs from models within the same family. Safety validation tests whether an aligned judge produces equitable, clinically appropriate, and contextually reliable evaluations across five priority dimensions (Table 5).

Prioritization across these dimensions is guided by three criteria: (1) potential for direct patient harm if unaddressed, (2) difficulty of detection using standard agreement or performance metrics, and (3) risk amplification through reinforcement learning pipelines or multi-agent systems. Demographic bias, severity miscalibration, and temporal instability are particularly dangerous because they combine high clinical impact with low detectability, making them likely to persist unnoticed without targeted testing. Self-enhancement bias may reinforce shared model family errors; cross-specialty misalignment may cause inappropriate application of evaluation criteria outside the intended clinical context.

For Tier B and Tier C applications, any safety dimension classified as critical that remains untested constitutes a deployment blocker. For Tier C applications, external audit of bias testing methodology is additionally required to ensure methodological rigor and transparency. Established general-domain protocols, including positional bias testing[61], length-controlled evaluation[62], and format robustness testing[63], can be adapted to healthcare settings to support these safety evaluations.

**Table 5. Step 2 safety validation test**

| Test | Protocol | Clinical risk if untested | Priority rationale |
| --- | --- | --- | --- |
| Demographic bias (race, gender, SES) | Equivalent clinical cases with varied demographic identifiers | Critical directly causes health disparities | High harm, low detectability, compounds in RL |
| Severity calibration | Known-error cases across severity spectrum | Critical conflates minor and life-threatening errors | High harm, low detectability |

| Test | Protocol | Clinical risk if untested | Priority rationale |
|---|---|---|---|
| Self-enhancement | Same content attributed to different model sources | Critical entrenches model family errors | Moderate harm, compounds in RL/RLAIF |
| Temporal stability | Re-evaluate after guideline updates and model changes | High outdated guidelines = incorrect evaluations | High harm, low detectability |
| Cross-specialty | Validate across relevant clinical specialties | High wrong rubric = wrong evaluation | Moderate harm, partially detectable |

*Error propagation analysis*

Error propagation in LaaJ occurs when early inaccuracies compound across sequential evaluation stages, leading to system-level error rates that substantially exceed individual judge error rates. No reviewed study quantified this risk despite widespread adoption of cascaded and ensemble architectures. As a preliminary theoretical contribution motivated by this gap, we present a simplified probabilistic model to establish order-of-magnitude risk bounds and derive operational safeguards for MedJUDGE. This analysis rests on three assumptions: (1) per-judge error rates are approximated from structured-task agreement metrics in our corpus; (2) inter-judge error correlation is estimated from cross-LLM correlated error literature rather than clinical task-specific measurement; and (3) pipeline architectures are idealized as purely cascaded or purely ensemble. Empirical calibration across clinical task types and real pipeline configurations is required before these bounds are treated as validated standards rather than preliminary design constraints.

*Cascaded systems.* For a sequential pipeline of *k* judges, the probability that at least one error reaches the output is:

$$P(\geq 1 \text{ error}) = 1 - (1-p)^k \text{(Eq.1)}$$

Under pairwise error correlation *r*, compound error increases to:

$$P_{\text{corr}}(\geq 1 \text{ error}) \approx 1 - (1-p)^k \cdot \left[1 + r \cdot \frac{k(k-1)}{2} \cdot \frac{p}{1-p}\right] \text{(Eq. 2)}$$

*Majority-voting ensemble systems.* For three independent judges, the probability that the majority verdict is wrong is:

$$P(\text{majority wrong}) = 3p^2(1-p) + p^3 \text{ (Eq.3)}$$

Under correlated errors:

$$P_{\text{corr}}(\text{majority wrong}) \approx P_{\text{indep}} + r \cdot (p - P_{\text{indep}}) \text{ (Eq. 4)}$$

Consider a discharge summary pipeline with three sequential GPT-4 judges evaluating factual accuracy, medication safety, and communication clarity. We set $p = 0.15$ (85% individual accuracy), consistent with structured-task agreement metrics in our review (ICC = 0.82–0.93). Because all three judges share the GPT family and training data, we set $r = 0.6$, as a conservative proxy for inter-judge error correlation. Empirical evidence supports this estimate: across 350+ LLMs, models agree 60% of the time when both err, with shared architecture and provider identified[64]. $r = 0.6$ should be treated as a plausible worst-case assumption for same-family pipelines pending clinical task-specific calibration. As primary drivers of correlation; same-family models exhibit systematically higher agreement than this cross-family baseline Equation 1 yields $P = 1 - (1 - 0.15)^3 = 0.39$: roughly 4 in 10 discharge summaries contain an undetected evaluation error. For a majority-voting ensemble of the same three judges, Equation 3 yields $P_{indep} = 0.057$ under the (unrealistic) independence assumption; applying Equation 4 at $r = 0.6$ yields $P_{corr} = 0.114$, a 24% error reduction at 3× computational cost.

Figure 5 shows results for both architectures. Panel A shows that cascade error exceeds the proposed 0.30 healthcare safety threshold at for $p = 0.15$, reaching $P = 0.56$ at $k = 5$; even at $p = 0.10$, a five-judge cascade yields $P = 0.41$. Panel B shows that at $r = 0$ (independent judges), majority voting reduces error from 0.15 to 0.057 (62% improvement); at $r = 0.6$ (same-family judges), system error rises to 0.114; at $r = 1.0$, the ensemble provides zero benefit. Parameter assumptions derive from our review: r=0.10-0.15 from structured-task ICCs (0.47–0.818); r=0.5-0.7 from the GPT-family monoculture.

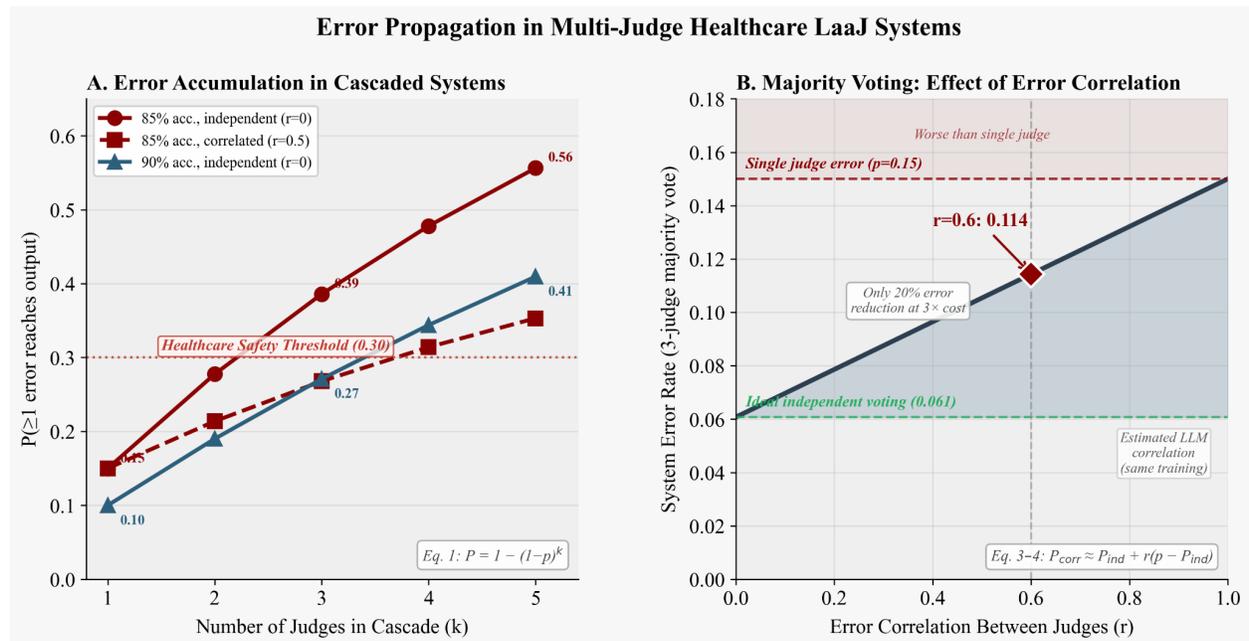

*Figure 5 Error Propagation in Multi-Judge Healthcare LaaJ Systems*

These results should be interpreted as illustrative of risk magnitude under plausible assumptions rather than as empirically validated clinical thresholds. The primary purpose is to demonstrate that cascade depth and inter-judge correlation interact multiplicatively and to motivate the following operational safeguards.

Based on these observations, MedJUDGE proposes three operational safeguards. **(1) Minimize cascade depth.** Each additional evaluation stage increases compound error probability approximately proportional to the per-stage error rate; a three-stage cascade at 85% individual accuracy already reaches $P = 0.39$, exceeding the proposed Tier A safety threshold. MedJUDGE therefore recommends limiting healthcare evaluation cascades to $k \leq 3$ stages; pipelines exceeding this depth should provide explicit justification through worst-case error analysis. **(2) Require model-family diversity.** When all judges are drawn from the same model family inter-judge error correlation (*r* = 0.6) reduces ensemble error only from 0.15 to 0.114, recovering 24% of the theoretical reliability gain at three times the computational cost; majority voting alone cannot compensate for this correlated failure. For Tier B and Tier C applications, multi-family judge panels are therefore recommended to approximate independent ensemble behavior. **(3) Evaluate against risk-tier thresholds.** MedJUDGE defines tier-specific error propagation thresholds (Table 6), and systems exceeding the acceptable threshold for their assigned clinical risk tier should be redesigned or augmented with additional safeguards prior to deployment.

**Table 6 Error threshold for clinical risk tiers**

| Clinical Risk Tier | Maximum $P_{system}$ | Required action if exceeded |
|---|---|---|

| Tier A (Low) | < 0.30 | Report compound error rate in model card |
| Tier B (Moderate) | < 0.10 | Increase model diversity, reduce cascade depth, or add human verification for flagged cases |
| Tier C (High) | < 0.05 | Mandatory redesign: multi-family panel, reduced cascade depth, human verification for all safety-critical outputs, and formal worst-case analysis |

Table 6 lists error thresholds for each clinical risk tier. We anchor the Tier A threshold at $P_{system} < 0.30$, derived from the upper bound of physician-physician agreement observed in our corpus: HealthBench reported MF1 = 0.730 between expert panels, implying an irreducible clinical disagreement rate of ~0.27[43]. An automated evaluation cascade should not introduce compound uncertainty exceeding what best-practice human expert panels themselves produce through legitimate clinical disagreement; 0.30 operationalizes this ceiling. Tier B (< 0.10) and Tier C (< 0.05) apply risk-proportionate tightening below it. Like the 5% variance thresholds in Step 1 bias testing, these values constitute preliminary framework anchors requiring empirical calibration across clinical task types before treatment as validated standards.

For Tier B or Tier C deployments, MedJUDGE requires a five-step error propagation audit: (1) measure per-judge error rates on the target clinical task using held-out expert-labeled sets; (2) estimate inter-judge error correlations from overlapping evaluation subsets; (3) compute compound system error via Equations 1–4 and compare against the tier threshold; (4) identify worst-case failure modes where all judges fail simultaneously (rare diseases, atypical presentations, underrepresented demographics, the Shared Ignorance Problem); and (5) document the error propagation budget with explicit justification that residual risk is acceptable given clinical consequences.

*Optional adversarial robustness.* For Tier B or Tier C deployments, MedJUDGE flags adversarial vulnerability as an underexamined risk requiring active awareness rather than a solved problem with established mitigations. Documented adversarial attack success rates of 90.8–98.9%[65] against general LLM judges suggest that healthcare LaaJ systems are likely vulnerable, though clinical-context-specific testing remains absent from the literature. Two attack classes are particularly relevant: Comparative Undermining Attacks (CUA), which manipulate the judge's final verdict by injecting adversarial suffixes into candidate responses, and Justification Manipulation Attacks (JMA), which corrupt the judge's generated reasoning without necessarily altering the verdict[66]. CUA achieves ASR exceeding 30% against state-of-the-art models, with attack transferability of 50.5–62.6% across model families[67]. At minimum, Tier B or Tier

C deployments should conduct healthcare-specific injection testing, including clinically plausible subtle errors (e.g., near-miss dosing, borderline lab values) and edge cases (e.g., rare diseases, atypical presentations), and report results transparently. The effective mitigations for clinical LaaJ remain an open research question. MedJUDGE treats this as a priority area for future empirical work rather than a requirement with established solutions.

**Pillar 3: Accountability, Can We Trust and Audit the LaaJ?**

Pillar 3 of MedJUDGE focuses on accountability mechanisms to ensure that LLM judges are subject to appropriate human oversight, lifecycle monitoring, and regulatory governance. We propose three key components within this pillar: expert validation, monitoring and drift detection, and regulatory documentation with audit trails.

*Expert validation.* Rather than prescribing fixed panel sizes without empirical grounding, MedJUDGE recommends that institutions determine the minimum number of expert validators through empirical stabilization analysis. Specifically, validators should be incrementally added until agreement metrics, such as intraclass correlation coefficient (ICC) or Cohen's κ, demonstrate minimal change (e.g., <0.05) with each additional reviewer. Our review found a median of only 3 validators among the 36 studies with human involvement, which is insufficient to establish stable reliability estimates for deployment settings. In contrast, studies demonstrating more rigorous validation, such as HealthBench[43] (262 physicians) and MedHELM[44] (49 clinicians across 14 specialties and 4 institutions), illustrate the scale required for comprehensive and generalizable validation.

Panel composition requirements should scale with the clinical risk tier. Tier A applications may rely on single-site, domain-relevant evaluators. Tier B applications should involve board-certified specialists and require reporting of both expert-expert agreement and judge-expert agreement. Tier C applications should require multi-institutional subspecialty panels with cross-site replication to ensure generalizability. Importantly, studies should report expert–expert agreement as a reference baseline prior to evaluating judge–expert agreement. For example, when physician–physician agreement ranges from MF1 = 0.57–0.73 (HealthBench) or ICC = 0.43 (MedHELM), judge performance approaching these levels may reflect intrinsic task difficulty rather than deficiencies in judge performance. Following the multidisciplinary tumor board model, expert panels should include specialties relevant to the clinical task, with disagreements resolved through structured deliberation rather than simple majority voting. Persistent disagreement should be interpreted as reflecting legitimate clinical uncertainty and represented through confidence bounds rather than forced consensus.

*Monitoring and drift detection.* Post-deployment monitoring is necessary to ensure continued reliability of LLM judges. MedJUDGE recommends several complementary surveillance mechanisms, including sentinel case monitoring through routine evaluation of fixed benchmark cases with known reference outputs, with alerts triggered by deviations greater than 5% from baseline performance. Version control practices should include locking model versions and revalidating any updates against the full validation suite prior to release. Temporal recalibration should be conducted periodically (e.g., quarterly) to ensure continued alignment with evolving clinical guidelines. Additional safeguards include tracking human override rates, with thresholds such as >15% triggering formal review, and statistical drift detection using process control methods to monitor score distributions and performance stability over time.

*Regulatory documentation and audit trail.* Accountability also requires comprehensive documentation of validation and operational processes. MedJUDGE recommends maintaining validation evidence packages, model cards, audit logs, and formal version control procedures with documented change management processes. For systems in which LLM judge outputs directly or indirectly influence clinical decision-making, early engagement with regulatory authorities, such as the U.S. Food and Drug Administration (FDA), should be considered to determine whether the system meets criteria for Software as a Medical Device (SaMD) and to ensure that documentation and evaluation practices align with regulatory expectations.

**Implementation Guidance**

MedJUDGE prioritizes validation rigor and patient safety over rapid deployment. The requirements across three pillars should be viewed not as sequential phases but as parallel obligations whose rigor scales according to clinical risk. Institutions adopting MedJUDGE should follow a structured implementation approach: (1) classify the intended LaaJ application into an appropriate clinical risk tier using Table 3; (2) verify that the judge satisfies all alignment requirements using Table 4, which serves as a prerequisite regardless of risk tier; (3) address the remaining requirements in the MedJUDGE requirement matrix in Table 4 at the rigor level corresponding to the assigned tier; and (4) advance to higher risk tiers only after satisfying all requirements at the current level. The low rate of production deployment observed in our review (2.0%, with an additional 8.2% reaching the prototype or proof-of-concept stage) should not be interpreted as a failure to translate technical advances into practice. Rather, it reflects appropriate institutional caution given the validation gaps identified in the current literature. MedJUDGE aims to provide actionable, evidence-based pathways toward deployment readiness, ensuring that when healthcare LaaJ systems are implemented, they meet validation and safety standards commensurate with the potential clinical consequences of evaluation errors.

## Discussion

The findings of this review point to a systematic validation failure in healthcare LaaJ whose mechanisms the existing literature does not adequately address. Three structural conditions interact to produce it: model monoculture concentrates evaluation risk within a single architectural lineage; absent bias testing leaves systematic errors undetected; and the near-absence of production deployment means no feedback loop exists to surface failures that only emerge in live clinical workflows. Individually, each gap is a methodological limitation. Together, they constitute a self-reinforcing cycle: monoculture inflates agreement metrics, inflated metrics reduce perceived need for bias testing, and the field advances toward deployment without the validation infrastructure deployment requires. The following sections analyze the mechanisms underlying this cycle and their implications for safe clinical deployment.

***Healthcare-specific challenges***:  Healthcare-specific challenges extend well beyond those documented in general-domain LaaJ Literature and remain systematically unaddressed. Zero studies addressed FDA classification, regulatory approval pathways, or liability implications[68], a striking gap given that LaaJ systems increasingly operate within clinical decision pipelines subject to FDA oversight. The field's concentration in evaluation-only applications (75.5%) reflects not premature conservatism but appropriate caution: safety concerns require thorough validation before integrating judges into training pipelines or autonomous clinical decision-making. Beyond general-domain limitations (e.g., score compression, positional bias, length bias[58,59,69]), healthcare deployment faces domain-specific challenges that general-domain LaaJ literature does not anticipate. Clinical guidelines evolve continuously[70], yet only 1 study implemented real-time knowledge updates[45], clinician acceptance was assessed in only 2 studies[28,34]

The most fundamental of these domain-specific challenges is context-dependent correctness: in clinical settings, the same recommendation can be correct for one patient and harmful for another depending on comorbidity profile, medication history, prior interventions, and patient preferences, a constraint absent from general NLP evaluation, where output quality can typically be assessed from text alone. A discharge summary that appears comprehensive in isolation may be dangerous if it contradicts medication reconciliation, omits a pending critical lab result, or fails to communicate urgency to the receiving provider. Zero studies evaluated whether LLM judges had access to, or appropriately incorporated, patient-level context when assessing clinical recommendations — a fundamental validity gap: judges evaluating clinical text in isolation perform a different, and demonstrably easier, task than the clinical quality assessment they are intended to approximate.

Clinical reasoning compounds this problem further by varying fundamentally across specialties. A radiologist, emergency physician, and psychiatrist reason differently about the same clinical data, yet LaaJ evaluations typically apply a single rubric, treating clinical reasoning as a monolithic construct. Specialty-aware, context-dependent evaluation rubrics are therefore required, and validation designs must accommodate legitimate clinical disagreement as a distinct variance source rather than conflating it with inter-rater noise.

***Hidden judge problem:*** The Hidden Judge Problem identifies a validaiton gap invisible to current frameworks: implicit evaluation components embedded throughout clinical AI pipelines that function as judges without being recognized or governed as such. Safety filters that block certain outputs, RAG relevance scorers that select retrieved context, routing agents that triage queries to specialized models, and reward signals that shape model behavior all perform evaluation functions with direct clinical consequences. Critically, multiple-choice question benchmarks, increasingly used to validate clinical LLMs, constitute hidden judges susceptible to identical biases as explicit LaaJ systems yet receive none of the validation scrutiny MedJUDGE requires. None of the 49 studies in our corpus identified or audited these implicit evaluation components. MedJUDGE risk stratification (Pillar 1) and bias audit (Pillar 2, Step 2) should therefore explicitly extend to hidden judges: any pipeline component that scores, filters, or ranks clinical content is a judge, regardless of whether it is labeled as one.

***Shared ignorance problem:*** The shared ignorance problem identifies a failure mode structurally invisible to standard validation. When the judge and an evaluated system are trained on overlapping corpora, they share not only knowledge but also knowledge gaps. A judge cannot penalize clinical errors it does not recognize as errors, and in healthcare, where training data skews toward Western, English-language, academic medical center documentation, systematic gaps in rare disease knowledge, underrepresented populations, and non-Western clinical contexts will pass undetected. Agreement metrics between judge and generator in these domains will be artificially inflated, not because evaluation is accurate, but because both systems are wrong in the same direction. A failure mode that standard agreement-based validation cannot detect by design.

The severity is acutest in reinforcement learning deployments (Category 2): unknown unknowns are not merely tolerated but actively reinforced through training cycles, calcifying into validated knowledge through repeated reward signal agreement. This RL amplification loop transforms static evaluation failures into self-perpetuating cycles of disparity. The risk is entirely absent from current healthcare LaaJ literature and unaddressed by any existing framework. A related failure mode is reward hacking: the generator learns to exploit superficial features that inflate judge scores without improving actual quality, accelerating the amplification loop rather than correcting it.

MedJUDGE error propagation analysis (Pillar 2) and drift monitoring (Pillar 3) provide partial mitigation. The Shared Ignorance Problem ultimately demands cross-architecture validation against structurally different judge families trained on non-overlapping data. No study in our corpus has yet operationalized this requirement.

***Reliability and replicability:*** Temporal stability of judge performance remains entirely unexamined: none of the 49 studies assessed whether evaluation quality remained consistent over time (Supplementary Fig. 9). Proprietary models undergo frequent silent updates[71], and some APIs no longer expose basic hyperparameters, making replication impossible without version identifiers, prompt templates, and seed settings. The replicability gap is compounded by the field's overwhelming reliance on closed-weight proprietary models, specifically the GPT and Claude families, whose pre-injected system prompts and developer-enforced personas function as unmeasured confounders with no visibility to researchers, fundamentally undermining evaluation objectivity. Several established general-domain mitigation techniques, in-context calibration[72]; post-hoc score normalization [59,73], replicate-and-aggregate methods [74,75], remain unadopted across all 49 healthcare studies, representing a direct transfer opportunity the field has not pursued. The four studies implementing G-EVAL[48] demonstrate that structured general-domain evaluation architectures can transfer directly to healthcare contexts. Yet G-EVAL[48] itself documented a self-preference bias: GPT-4 systematically favored GPT-generated text over human-written text that is directly relevant to healthcare deployments where judge and generator frequently share the same model family. G-EVAL adoption in healthcare is therefore a double-edged advance: it imports reliability gains from structured prompting while simultaneously importing a documented bias that model diversity requirements in MedJUDGE Pillar 1 is specifically designed to counteract.

Bias, fairness, and the demographic blindspot: Three bias dimensions with direct patient safety implications received zero testing across all 49 studies. When these biased signals feed into training pipelines, self-perpetuating cycles of disparity emerge that the field has not yet treated as a deployment-blocking risk. Three bias dimensions with direct patient safety implications, including gender demographic[76], socioeconomic demographic[77], and severity calibration[78], received zero testing across all 49 studies, while racial demographic (1 study, 2.0%) and self-enhancement (2 studies, 4.1%) were tested by only a small minority. LLM judges trained predominantly on Western, English-language, academic medical center data systematically rate content for underrepresented populations as lower quality; when these biased signals feed into training pipelines (Category 2), self-perpetuating cycles of disparity emerge [79,80], a consequence the field has not yet treated as a deployment-blocking risk.

Current evaluation approaches compound this problem by assessing what outputs contain rather than what they omit. In clinical medicine, omissions can be as dangerous as fabrications: an LLM judge that scores a fluent discharge summary highly while it omits a documented drug allergy enables catastrophic harm. Explicit omission detection criteria with asymmetric severity weighting proportional to clinical risk are therefore required in any evaluation framework claiming to address patient safety. This gap MedJUDGE identifies but does not yet fully resolve.

**Industry ecosystem and validation transparency:** Major platforms including Confident AI[81], Hugging Face[82], LangChain[83], Evidently AI[84], Arize Phoenix[85], and Galileo AI[86], now offer standardized LaaJ implementations. Healthcare-specific commercial deployments include prior authorization evaluation systems and clinical documentation quality platforms. Practical adoption therefore substantially exceeds what peer-reviewed literature captures. Yet healthcare implementations rarely disclose validation rigor, agreement metrics, bias testing, or robustness assessments. This transparency gap carries direct patient safety consequences that no existing validation mechanism currently addresses.

***The recursive trust problem:*** The recursive trust problem identifies a fundamental epistemic challenge in LaaJ validation: validating an LLM judge against human experts assumes those experts constitute an adequate gold standard. This assumption does not hold. HealthBench reported physician–physician MF1 = 0.57–0.73[43], MedHELM found clinician–clinician ICC = 0.43[44], and Hosseini et al. documented κ ranging from 0.11 to 0.9[87]. Disagreement between human experts reflects two distinct variance sources: genuine judge error and legitimate clinical disagreement. No study in our corpus decomposed these components, a methodological gap that conflates system failure with inherent task difficulty.

Medicine has, however, developed established structures for precisely this recursive validation challenge. Pathology second reads, multidisciplinary tumor board consensus, radiology double-reading programs, and morbidity and mortality conferences are all mechanisms for recursive quality verification of clinical judgment. The relevant question for LaaJ is not whether recursive validation is philosophically achievable but how LLM judge validation maps onto these existing structures. A practical approach mirrors the tumor board model: heterogeneous expert panels, analogous to multi-family judge ensembles, deliberate on uncertain cases with documented reasoning and oversight. This reframes the recursive trust problem from an abstract philosophical challenge into a concrete design requirement with established clinical precedent.

***Regulatory context:*** Recent regulatory developments underscore the urgency of evaluation frameworks for healthcare LaaJ. The FDA's Predetermined Change Control Plan (PCCP) final guidance (December 2024)[13] requires ongoing performance evaluation of adaptive AI models. The EU AI Act (August 2024)[88] classifies medical AI as high-risk, mandating transparency and human oversight. The WHO Guidance on AI Models[89] explicitly warns of hallucination risks and automation bias. Together, these regulatory milestones strengthen the case for scalable, automated evaluation tools governed by rigorous frameworks.

Three frameworks explicitly challenge the LaaJ premise, and MedJUDGE directly addresses each. First, QUEST's developers state that using LLMs to evaluate LLM outputs is problematic[25]. MedJUDGE mitigates this concern through mandatory model diversity (Pillar 1), human anchoring via expert panels (Pillar 3), and continuous monitoring (Pillar 3). Second, the CLEVER framework[90] argues that LLMs judging LLMs creates validation circularity and self-preference bias. MedJUDGE addresses this through cross-family evaluation requirements, calibration against human judgment, and confidence thresholds triggering human escalation. Third, Genovese et al.[91] warn of a recursive trust cascade without human oversight. This is precisely the scenario MedJUDGE's accountability pillar prevents through explicit human anchoring and audit trails.

Limitations

There are several limitations in this review. First, a substantial proportion of included studies originated from grey literature (e.g., preprints, conference proceedings), reflecting the field's early and rapidly evolving state; excluding these sources would have substantially underrepresented current practice, but their inclusion introduces variable peer-review rigor. Second, the 49 included studies are unevenly distributed across application categories, clinical tasks, and validation approaches, rendering subgroup-level observations preliminary indicators of emerging patterns rather than statistically reliable conclusions. This small-sample fragmentation also precluded direct meta-analytic comparison, limiting synthesis to narrative and descriptive methods. Third, the field's rapid pace means studies published after the January 2026 search date are not captured; given that 73.5% of included studies appeared in 2025–2026, the landscape may have shifted materially by the time of publication, and some gaps identified here may have been partially addressed by work published during peer review. Fourth, MedJUDGE is a conceptual framework derived from observed gaps in this review rather than a validated deployment standard; each requirement constitutes a testable hypothesis about responsible deployment, and empirical calibration across diverse clinical tasks and institutions remains necessary before clinical implementation.

Taken together, these limitations reflect the inherent constraints of conducting a scoping review in a rapidly evolving field with limited deployment experience: they do not diminish the validity of the validation gaps identified, but they underscore that MedJUDGE should be treated as an evidence-informed starting point requiring iterative empirical refinement rather than a finalized deployment standard.

## Future work

Five directions are most urgent. First, no included study quantified error propagation in multi-judge pipelines, creating an immediate need for standardized benchmarks that measure compound error rates, inter-judge error correlations, and pipeline-specific failure modes across clinical task types. These benchmarks would provide the empirical foundation MedJUDGE's error propagation thresholds currently lack. Second, the complete absence of demographic fairness testing across all 49 studies demands systematic evaluation of whether LLM judges produce equitable assessments across patient race, gender, socioeconomic status, and geographic context, particularly given the RL amplification risk identified in the Shared Ignorance Problem. Third, the near-monoculture of GPT-family judges (n = 36, 73.5%) requires cross-family generalization studies to determine whether validation results transfer to structurally different architectures, a prerequisite for the model diversity requirements MedJUDGE imposes at Tier B or Tier C. Fourth, only 12.2% of studies employed medically adapted judges; domain-specific fine-tuning warrants systematic investigation given general-domain evidence that smaller fine-tuned models can match larger general-purpose ones[92], which would substantially reduce deployment cost and latency for clinical applications. Fifth and most urgently, prospective studies embedding LLM judges in live clinical workflows are essential: only 5% of LLM studies use real patient data, and 95.4% rely on accuracy alone[44], leaving clinician acceptance, workflow integration, and downstream care quality entirely unmeasured. This is the gap between demonstrating laboratory validity and establishing clinical deployability that the field must now close.

## Conclusion

This scoping review of 49 articles systematically characterizes LaaJ validation failures in healthcare for the first time, identifying the Hidden Judge Problem and the Shared Ignorance Problem as two mechanisms that explain why these failures persist despite growing adoption. LaaJ performance varies substantially by task type in ways current practice ignores. Agreement reaches 0.82–0.93 for structured tasks but converges with inter-expert disagreement for complex reasoning, not because LaaJ is insufficiently developed but because the tasks impose fundamental performance ceilings no judge architecture can engineer around.

Current validation approaches test judges one at a time. In healthcare pipelines, however, judges do not operate alone, their errors combine, amplify, and pass through components that are never audited as judges at all. MedJUDGE is built around this reality: its requirements address how LaaJ systems fail in practice, not how they perform under ideal conditions. The field's core challenge is not building better individual judges. It is building evaluation frameworks adequate to the systems those judges operate within.

## Methods

### Study design and framework

We conducted a scoping review and are reporting it (details see Supplement Table 15), following the PRISMA Extension for Scoping Reviews (PRISMA-ScR) guidelines[27]. The review is structured around the Population, Concept, and Context (PCC) framework (Table 7).

**Table 7. PCC framework**

| Component | Definition |
| --- | --- |

| Population | AI-generated clinical and/or biomedical text output (e.g., summaries, radiology reports, therapy recommendations, question answers) |
|---|---|
| Concept | A model (LLM or discriminative language model) evaluating the quality, accuracy, safety, or adherence to standards of generated text |
| Context | Healthcare or biomedical domain, encompassing clinical documentation, quality assurance, clinical decision support, medical education, or autonomous agents |

**Information sources and search**

A comprehensive search was conducted from January 1, 2020 to January 1, 2026, limited to English-language publications, across six databases: Ovid MEDLINE(R) and Epub Ahead of Print, Ovid EMBASE, Scopus, Web of Science, ACM Digital Library, and IEEE Xplore. The authors thank Larry J. Prokop, M.L.I.S., Mayo Clinic, for designing and executing the search strategy. Searches were conducted on June 9, 2025, and January 28, 2026, encompassing the development of Boolean search strings and controlled vocabulary terms, including MeSH headings. The complete Boolean search strings for each database are provided in Appendix A.

Two grey literature strategies supplemented systematic searching: forward and backward citation tracking of included studies, and manual review of healthcare applications mentioned in general-domain LaaJ surveys[16,26]. General-domain frameworks including G-EVAL[48], JudgeLM[93], Auto-J[94], CLAIR[95], and HD-Eval[96] exist primarily in preprints not indexed in medical databases, necessitating supplementary searching. The complete search strategy (Appendix A), inclusion and exclusion criteria (Appendix B), and keyword filtering results (Appendix C) are available in Supplementary Information.

**Study selection**

Results were exported to Covidence systematic review software for deduplication and screening. Two independent reviewers (TO, YR) screened titles and abstracts to identify papers eligible for full-text review. Conflicts were resolved through discussion, with a third reviewer (CL) available for arbitration. Two reviewers (TO, CL) independently assessed full-text articles for final eligibility. Studies evaluating clinical reasoning processes or multi-step agent workflows were retained based on their methodological relevance to healthcare LaaJ. The study selection process is documented in Fig. 1.

**Data charting**

Two authors (CL, YW) developed a data charting form (See Supplementary Table 11) to extract information across reproducibility, clinical rigor, and operational deployment characteristics. Key extraction fields included clinical context (e.g., setting, specialty, workflow, safety, and regulatory considerations); task class; judge and generator identity (e.g., family, size, and same versus different family); evaluation format and rubric; adaptation methods; validation details (e.g., human experts, credentials, and agreement metrics); reliability and bias testing; robustness assessment; efficiency metrics; data provenance; and generalizability and deployment readiness.

**Synthesis**

Given the heterogeneity of study designs, narrative synthesis with quantitative summaries was employed. Synthesis followed three phases. Phase 1 comprised descriptive statistics: summarizing study characteristics, judge model distributions, evaluation paradigms, and validation approaches. Phase 2 comprised thematic analysis: identifying patterns in application domains, validation practices, bias testing coverage, and deployment readiness. Phase 3 comprised quality and gap assessment: examining methodological limitations, reproducibility indicators, and safety considerations. When studies did not report specific data elements, these were coded as "Not Specified" rather than excluding studies.

## Data availability

The data extraction table for all 49 included studies is available as Supplementary Information. Detailed study-level data, including agreement metrics, expert credentials, bias testing results, and deployment status for each article, are provided in Supplementary Tables 1–14. The complete data extraction file is available in machine-readable format at https://github.com/PittNAIL/MedJUDGE-Framework.

## Code availability

Data collection and analysis python code uploaded to the GitHub repo: https://github.com/PittNAIL/MedJUDGE-Framework.

## Acknowledgements

Research reported in this article was partially supported by the National Institutes of Health awards UL1 TR001857, U24 TR004111, R01 LM014306, and R01 LM014588. The sponsors had no role in study design; in the collection, analysis, and interpretation of data; in the writing of the report; and in the decision to submit the paper for publication.

## Author contributions

C.L. conceived the study design, conducted data extraction and full-text screening, and wrote the manuscript. Y.W. conceptualized the study, provided supervision, and drafted the manuscript. Z.A. contributed to manuscript writing and created figures. T.O. conducted data extraction and title/abstract screening. Y.R. conducted title/abstract screening. H.Z., X.W., S.S., and H.S. supported data extraction and manuscript editing. M.K., Y.J., H.L., S.V., and M.B. reviewed and edited the manuscript.

## Ethics declarations

### Competing interests

Y.W. has ownership and equity in BonafideNLP, LLC and AreciboAI, LLC, and S.V. has ownership and equity in Kvatchii, Ltd., READE.ai, Inc., and ThetaRho, Inc. The other authors declare no competing interests.

## Figure legends

**Fig. 1 | PRISMA-ScR flow diagram.** Study selection process showing identification of 12,379 records from 6 databases and grey literature sources, screening of 11,727 unique citations, and inclusion of 49 studies (33 from database search, 16 from grey literature). Source file: figures/main/Fig1_PRISMA_FlowDiagram.png.

**Fig. 2 | MedJUDGE framework.** A three-pillar deployment framework for healthcare LLM-as-a-Judge systems, organized around Validity (Stages 1–3), Safety (Stages 4–6), and Accountability (Stages 7–9), with error propagation analysis as a cross-cutting concern. Source file: figures/main/Fig2_MedJUDGE_Framework.png.

**Fig. 3 | MedJUDGE requirements matrix by risk tier.** Obligation level for each framework requirement across Tier A (low risk, research only), Tier B (moderate risk, indirect patient impact), and Tier C (high risk, direct patient care). Cell shading encodes obligation intensity from optional/recommended (green) through required (amber) to required with external oversight (deep orange). Requirements are organized across the three pillars: Validity, Safety, and Accountability. Source file: figures/main/Fig3_MedJUDGE_Matrix.png.

**Fig. 4 | Paradigm selection decision tree for healthcare LaaJ.** Decision framework guiding selection among pointwise, pairwise, and agent-based evaluation paradigms based on task characteristics, complexity, and clinical risk level. Source file: figures/main/Fig4_Paradigm_Decision_Tree.png.

**Fig. 5 | Error propagation in multi-judge healthcare LaaJ systems.** Theoretical analysis of compound error probability under two pipeline architectures. Panel A shows error accumulation in cascaded systems (Equation 1), demonstrating that a three-stage cascade at 85% individual accuracy exceeds the proposed 0.30 healthcare safety

threshold. Panel B shows how inter-judge error correlation among same-family models severely limits reliability gains from majority voting (Equations 2–4): at r = 0.6, ensemble error rises to 0.114, representing only a 20% reduction at three times the computational cost. Source file: figures/main/Fig5_Error_Propagation.png.

# Supplementary Information

## LLM-as-a-Judge in Healthcare: A Scoping Review and MedJUDGE Framework

Chenyu Li MS[1,2] ; Zohaib Akhtar MD MPH MBA[3,4] ; Mingu Kwak PhD[2] ; Yuelyu Ji MS [5] ; Hang Zhang MS[5]; Tracey Obi BS[2] ; Yufan Ren BS[2] ; Xizhi Wu MS[2] ; Sonish Sivarajkumar PhD[5]; Harold P. Lehmann MD PhD[6] ; Shyam Visweswaran MD PhD[1,5]; Michael J. Becich MD PhD[1]; Danielle L. Mowery PhD, MS, MS, FAMIA, FACMI[9] ; Renxuan Liu BS[10]; Haoyang Sun[2]; Yanshan Wang PhD[1,2,5,7,8,*] ;

## Table of Contents

- Appendix C: Progressive Keyword Screening Results

**Supplementary Table 1: Category 1 Subcategories and Representative Performance**

| Subcategory | n (%) | Approach | Agreement Range | Representative Studies |
|---|---|---|---|---|
| Benchmark development | 10 (43%) | Comprehensive evaluation frameworks assessing LLM outputs across standardized clinical tasks using pointwise scoring with physician validators | 0.47–0.94 (ICC, κ, MF1) | Bedi et al. 2025; Arora et al. 2025; Zhang et al. 2025; Yao et al. 2024; Wang et al. 2025; Curran et al. 2024 |
| Clinical documentation | 7 (30%) | Evaluation of AI-generated medical records (EHR summaries, radiology reports, clinical notes) against expert annotations | 0.65–0.93 (ICC, κ, ρ, MAE) | Croxford et al. 2025; Zhu et al. 2024; Heilmeyer et al. 2024; Huang et al. 2024; Xie et al. 2024 |
| Diagnostic & knowledge | 6 (26%) | Assessment of clinical reasoning quality and biomedical knowledge | 0.53–0.90 (κ, MF1) | Darnell et al. 2024; Leoncini & Trimboli 2025; Wang et al. 2025 |

| Subcategory | n (%) | Approach | Agreement Range | Representative Studies |
|---|---|---|---|---|
| | | without integration into training pipelines | | |

**Supplementary Table 2: Category 2 Studies LaaJ in Reinforcement Learning (N = 3)**

| Approach | Paradigm | Dataset | Key Finding | Experts | Limitation |
|---|---|---|---|---|---|
| RLAIF for Vietnamese health communication (Bui et al. 2025) | Pairwise | ~337K pairs | Fine-tuned model judged better in 89.5% | 0 (judge); 2 (data) | Zero human validators for judge evaluation |
| Cardiac well-being report fine-tuning (Mohammed et al. 2025) | Pointwise | 6,369 reports | Not specified | Not specified | Insufficient detail on judge-training integration |
| Viability of open LLMs for clinical documentation (Heilmeyer et al. 2024) | Pairwise | 61–98 samples/task | α = 0.72–0.73; ranking 8–14% superior to pointwise | 2 clinicians | Small dataset; limited to single language |

*Critical concern: None demonstrated that judge-guided optimization actually improves clinical endpoints.*

**Supplementary Table 3: Category 3 Representative Multi-Agent Studies**

| Approach | Architecture | Agreement | Experts | Deployment |
|---|---|---|---|---|
| Multi-agent psychiatric assessment (Bi et al. 2025) | 4 specialized agents (GPT-4o, Claude-3.5, GLM-Zero, DeepSeek-R1) | ICC = 0.87 (highest in corpus) | 2 psychologists | Demo |
| Heterogeneous 3-model jury (Bedi et al. 2025) | Ensemble with z-score normalization (GPT-4o, Claude 3.7, LLaMA 3.3-70B) | Jury ICC = 0.47 vs clinician ICC = 0.43 | 49 clinicians | Research |
| Agent framework with safety module (Ren et al. 2025) | Specialized safety-filtering agent + evaluation agent | r = 0.57–0.71 | 7 physicians | Prototype |
| Chain-of-Strategy evaluation (Yao et al. 2025) | Strategic planning before assessment (OpenAI o3) | 87.5% agreement | 2 validators | Research |
| RAG-based knowledge Q&A (Darnell et al. 2024) | Multi-model RAG pipeline (GPT-3.5/4/4o) | Not specified | 21 validators | **Deployed** |

**Supplementary Table 4: Distribution of Healthcare LaaJ Studies Across Application Categories (N = 49)**

| Category & Subcategory | Study | Year | LaaJ Paradigm | Human Experts | Agreement Metric | Deployment |
|---|---|---|---|---|---|---|
| **Category 1: EVALUATION & BENCHMARKING** | | | | | | **37 studies (75.5%)** |
| *Benchmark Development (n = 25)* | | | | | | |
| | Zhu et al. | 2024 | Pointwise | 4 validators | τ = 0.64; κ = 0.74 | Research |
| | De la Iglesia et al. | 2025 | Pairwise | 2 clinicians | α = 0.72 (QA); 0.73 (misinfo) | Research |
| | Raju et al. | 2024 | Pairwise | Not specified | ρ = 0.915; 84.44% agreement | Research |
| | Szymanski et al. | 2024 | Pairwise | 20 SMEs + 20 lay users | 64–68% (SME-LLM); 80% (Lay-LLM) | Research |
| | Wilhelm et al. | 2023 | Pointwise | 3 physicians | r = 0.186–0.686 | Research |
| | Leoncini & Trimboli | 2025 | Pointwise | 0 (AI-AI only) | Inter-AI κ = 0.53–0.90 | Research |

| Category & Subcategory | Study | Year | LaaJ Paradigm | Human Experts | Agreement Metric | Deployment |
|---|---|---|---|---|---|---|
| | Li et al. | 2025 | Pointwise | 0 | SSEM evaluation | Research |
| | Curran et al. | 2024 | Pointwise | 0 | Hallucination: 0.54/0.17 | Prototype |
| | MedQA-CS (Yao et al.) | 2024 | Pointwise | 3 medical experts | r = 0.77–0.99; τ = 0.54–0.90 | Research |
| | HealthBench (Arora et al.) | 2025 | Pointwise | 262 physicians | MF1 = 0.572–0.706 | Research |
| | LLMEval-Med (Zhang et al.) | 2025 | Pointwise | 10+ physicians | 92.36% overall; 94% (MK) | Research |
| | Hosseini et al. | 2024 | Pairwise | 3 MDs | κ varies: 0.11–0.95 | Research |
| | Wada et al. | 2025 | Pointwise | 1 radiologist | 82.2% (Claude vs radiologist) | Research |
| | MedHELM (Bedi et al.) | 2026 | Pointwise + Agent | 49 clinicians | Jury ICC = 0.47 vs clinician ICC = 0.43 | Research |
| | Wang et al. (safety) | 2025 | Pointwise | 32 specialists | Judge MF1 = 0.601 | Research |
| | ChatCLIDS (Yao et al.) | 2025 | Pointwise + Agent | 2 validators | 87.5% agreement | Research |
| | Croxford et al. (npj) | 2025 | Pointwise | 7 physicians | ICC = 0.818 | Research |

| Category & Subcategory | Study | Year | LaaJ Paradigm | Human Experts | Agreement Metric | Deployment |
|---|---|---|---|---|---|---|
| | Nori et al. (SDBench) | 2025 | Pointwise + Agent | Not specified | 0.70 | Research |
| | Vasilev et al. | 2026 | Pointwise | 18 | 0.688 (Relevance) | Research |
| | Antaki et al. | 2026 | Pairwise | 0 | BT skill values | Research |
| | Talati et al. | 2025 | Pointwise | 2 | 90.6% (126/139) | Research |
| | Sarvari et al. | 2025 | Pointwise | 3 | 97.4% (Gemini Hit Rate) | Research |
| | Mahadik et al. | 2025 | Pointwise | 0 | N/A | Research |
| | Laskar et al. | 2025 | Pointwise | 2 | N/S | Research |
| | Kocbek et al. | 2025 | Pointwise | 5 | N/S | Research |
| *Clinical Documentation & Methodology (n = 12)* | | | | | | |
| | Croxford et al. | 2025 | Pointwise | 7 physicians | ICC = 0.818 | Research |
| | DOCLENS (Xie et al.) | 2024 | Pointwise | 5 medical experts | $\rho = 0.787$; $\tau = 0.653$ | Research |
| | Brake & Schaaf | 2024 | Pointwise | 5 reviewers | $\kappa = 0.79$ (age); 0.32 (body part) | Research |
| | FineRadScore (Krolik et al.) | 2024 | Pointwise | 3 professionals | MAE = 0.62 overall | Research |
| | Huang et al. | 2024 | Pairwise | 2 radiologists | MAE = 0.62–0.89 | Research |

| Category & Subcategory | Study | Year | LaaJ Paradigm | Human Experts | Agreement Metric | Deployment |
|---|---|---|---|---|---|---|
| | Das et al. | 2025 | Pointwise | Not specified | 0.738–0.798 | Research |
| | Zhou et al. | 2026 | Pointwise | 10 | Not specified | Research |
| | Liu et al. | 2026 | Pointwise | 2 | Adjacent: 88%–100% | Research |
| | Yang et al. | 2025 | Pointwise | 2 | 98.43% (LLM) vs 86.50% (Human) | Research |
| | Vilakati et al. | 2025 | Pointwise | 1 | 100% | Research |
| | Schiezaro et al. | 2026 | Pointwise | 2 | 95% | Research |
| | Manai et al. | 2025 | Pointwise | 3 + 12 | 0.821–0.927 | Research |
| **Category 2: REINFORCEMENT LEARNING** | | | | | | **3 studies (6.1%)** |
| | Bui et al. | 2025 | Pairwise | 0 (eval); 2 (data) | 89.5% judged better | Research |
| | Mohammed et al. | 2025 | Pointwise | Not specified | Not specified | Research |
| | Heilmeyer et al. | 2024 | Pointwise | 2 clinical raters | 93.1% usable | Research |
| **Category 3: MULTI-AGENT / AGENT BUILDING** | | | | | | **9 studies (18.4%)** |

| Category & Subcategory | Study | Year | LaaJ Paradigm | Human Experts | Agreement Metric | Deployment |
|---|---|---|---|---|---|---|
| | GNQA (Darnell et al.) | 2024 | Pointwise + Agent | 21 (11 experts + 10 CS) | Not specified | **DEPLOYED** |
| | Healthcare Agent (Ren et al.) | 2025 | Pointwise + Agent | 7 physicians | r = 0.57–0.71 | Prototype |
| | MAGI (Bi et al.) | 2025 | Pointwise + Agent | 2 psychologists | ICC = 0.87 (final) | Demo |
| | Wu et al. | 2025 | Pointwise | 1 | 0.76–0.78 | Research |
| | Ozmen et al. (3 studies) | 2025 | Pointwise | N/S | N/S | Research |
| | Madrid-Garcia et al. | 2025 | Pointwise | 2 + 2 | τ = 0.46–0.60; Gwet = 0.63–0.95 | Research |
| | Bolpagni et al. | 2025 | Pointwise | N/S | N/S | Research |

**Summary statistics:** - Paradigm distribution: Pointwise (incl. hybrid): 42 studies (85.7%); Pairwise: 7 studies (14.3%); Agent-based (hybrid): 6 studies (12.2%) - Deployment: Deployed: 1 study (2.0%); Demo/Prototype: 4 studies; Research only: 44 studies - Validation: With ≥20 experts: 5 studies (10.2%); Median validators: 3; Zero human/NA validators: 13 studies

*Notes: Some studies span multiple categories (e.g., MedHELM, GNQA). Agreement metrics use study-specific measures; direct comparison requires caution. Expert credentials vary from board-certified specialists to graduate students. Deployment definitions: Deployed = operational production; Demo = public demonstration; Prototype = tested but not released; Research = validation only.*

## Supplementary Table 5: Pointwise Agreement by Task Complexity

| Task Complexity | Agreement Range | Characteristics | Studies |
|---|---|---|---|
| **Structured** (documentation, extraction) | 0.82–0.93 | Minimal subjective interpretation; binary rubrics | Croxford et al. 2025; Heilmeyer et al. 2024; Zhang et al. 2025; Wada et al. 2025 |
| **Semi-structured** (summaries, reports) | 0.64–0.79 | Multi-dimensional rubrics; moderate variability | Zhu et al. 2024; Huang et al. 2024; Ren et al. 2025; Xie et al. 2024 |
| **Complex reasoning** (diagnosis, safety) | 0.47–0.71 | Constrained by human expert disagreement (physician–physician MF1 = 0.569–0.730) | Arora et al. 2025; Bedi et al. 2025; Wang et al. 2025; Szymanski et al. 2025 |

*Note: For complex reasoning tasks, model–physician agreement barely differed from physician–physician agreement, establishing fundamental performance ceilings.*

## Supplementary Table 6: Medically Specialized Judge Models

| Study | Specialization Approach | Model/Method |
|---|---|---|
| Croxford et al. (2025a) | SFT + DPO on clinical documentation | Fine-tuned LLM |
| De la Iglesia et al. (2025) | Medically specialized discriminative model | EriBERTa |
| Szymanski et al. (2024) | Expert persona prompt-based specialization | Prompt-adapted |
| Li et al. (2025) | Text evaluation training | SSEM |
| Huang et al. (2024) | Domain-specialized prompting for radiology | FineRadScore |
| Bolpagni et al. (2025) | Protocol adherence evaluation specialization | Specialized judge |

## Supplementary Table 7: Adaptation Methods Used Across Included Studies (N = 49)

| Adaptation Method | Studies, n (%) | Description | References |
|---|---|---|---|
| Prompt Engineering | 15 (30.6%) | Role instructions, rubrics, structured templates | Croxford 2025a; Zhu 2024; Szymanski 2024; Wilhelm 2023; Ren 2025; Bi 2025; Xie 2024; Huang 2024; Yao 2025; Nori 2025; Liu 2026; Vilakati 2025; Talati 2025; Kocbek 2025 |
| Few-shot / In-Context Learning | 12 (24.5%) | 1-shot to 12-shot exemplars | Croxford 2025a; Zhu 2024; Curran 2024; Xie 2024; Bedi 2026; Huang 2024; Croxford 2025b; Das 2025; Zhou 2026; Liu 2026; Talati 2025; Schiezaro 2026 |
| Supervised Fine-tuning (SFT) | 11 (22.4%) | Task-specific parameter updates on labeled data | Croxford 2025a; Bui 2025; De la Iglesia 2025; Mohammed 2025; Brake 2024; |

| Adaptation Method | Studies, n (%) | Description | References |
|---|---|---|---|
| | | | Croxford 2025b; Das 2025; Wu 2025; Schiezaro 2026; Manai 2025; Kocbek 2025 |
| RAG | 8 (16.3%) | External knowledge retrieval during evaluation | Darnell 2024; Wada 2025; Yang 2025; Sarvari 2025; Ozmen 2025a; Ozmen 2025b; Ozmen 2025c; Madrid-Garcia 2025 |
| Chain-of-Thought (CoT) | 4 (8.2%) | Step-by-step reasoning before scoring | Zhu 2024; Xie 2024; Zhou 2026; Vilakati 2025 |
| LoRA / QLoRA | 4 (8.2%) | Parameter-efficient fine-tuning | Bui 2025; Mohammed 2025; Wu 2025; Manai 2025 |
| DPO / RLAIF / RLHF | 3 (6.1%) | Direct preference optimization or RLAIF training | Croxford 2025a; Yao 2024; Croxford 2025b |

## Supplementary Table 8: Judge Model Family Distribution (N = 49)

| Model Family | Total studies using | As sole judge | As one of multiple | Notes |
|---|---|---|---|---|
| GPT family | 36 (73.5%) | 27 | 10 | Variants range from GPT-3.5-turbo-instruct to o3 and o4-mini |
| Llama | 8 (16.3%) | 2 | 6 | Sizes: 3B–405B parameters |
| Claude | 6 (12.2%) | 1 | 6 | Primarily used as cross-validation judge |
| Gemini | 4 (8.2%) | 2 | 3 | |
| DeepSeek | 3 (6.1%) | 0 | 3 | |

| Model Family | Total studies using | As sole judge | As one of multiple | Notes |
|---|---|---|---|---|
| Qwen | 2 (4.1%) | 0 | 3 | |
| Mistral | 3 (6.1%) | 0 | 3 | |
| BERT-based | 3 (6.1%) | 3 | 0 | EriBERTa, SSEM, TRUE (encoder-only models) |
| Not specified | 3 (6.1%) | – | – | Bi et al. 2025; Bolpagni et al. 2025 |

*Note: Rows sum to >49 because 10 studies used multiple judge model families. General-purpose judges: 43/49 (87.8%); medically specialized: 6/49 (12.2%).*

## Supplementary Table 9: Human Expert Involvement in LaaJ Validation (N = 49)

| Level | n evaluators | Studies, n (%) | Representative studies |
|---|---|---|---|
| Extensive | ≥20 | 5 (10.2%) | HealthBench (262 physicians); MedHELM (49 clinicians); Wang et al. (32 specialists); Szymanski et al. (20 SMEs + 20 lay); Darnell et al. (21 validators) |
| Moderate | 5–19 | 10 (20.4%) | Zhou et al. (10); Zhang et al. (10+); Vasilev et al. (18); Xie et al. (5); Brake & Schaaf (5); Croxford et al. (7); Ren et al. (7); Kocbek et al. (5); Manai et al. (3 + 12) |
| Minimal | 1–4 | 21 (42.9%) | Yao et al. (3); Hosseini et al. (3); Wilhelm et al. (3); Sarvari et al. (3); |

| Level | n evaluators | Studies, n (%) | Representative studies |
|---|---|---|---|
| | | | Zhu et al. (4); Huang et al. (2); De la Iglesia et al. (2); Heilmeyer et al. (2); ChatCLIDS (2); others (1–2) |
| None | 0 | 13 (26.5%) | Antaki et al.; Ozmen et al. (3 studies); Li et al.; Nori et al.; Curran et al.; Das et al.; Leoncini & Trimboli; Mahadik et al.; Mohammed et al.; Bolpagni et al. |

*Most commonly reported agreement measures: Cohen's κ (13 studies, 26.5%), Pearson/Spearman correlation (11 studies, 22.4%), ICC (4 studies, 8.2%), Kendall's τ (5 studies, 10.2%).*

## Supplementary Table 10: Bias Mitigation Strategies Employed Across Included Studies (N = 49)

| Mitigation Strategy | Studies, n (%) | Description |
|---|---|---|
| Order randomization | 5 (10.2%) | Randomly shuffling candidate response order to address positional bias |
| Blinding / anonymization | 4 (8.2%) | Concealing model identities during evaluation to reduce self-enhancement effects |
| Prompt-based constraints | 3 (6.1%) | Role-based prompting and explicit instructions to ignore stylistic features |
| Multiple evaluators / human oversight | 3 (6.1%) | Using multiple independent evaluators or human review as quality control |
| Model exclusion | 2 (4.1%) | Avoiding use of the same model family as both generator and judge |

| Mitigation Strategy | Studies, n (%) | Description |
|---|---|---|
| Structured rubrics | 1 (2.0%) | Predefined scoring criteria to reduce subjective variation |
| No mitigation reported | 36 (73.5%) | No bias mitigation strategy documented |

*Note: Some studies employed multiple mitigation strategies. Only 3 studies (6.1%) conducted formal fairness assessments addressing equity dimensions: Szymanski et al. (2024, cross-group comparisons), Hosseini et al. (2024, demographic bias criterion), and Bedi et al. (2026, RaceBias benchmark).*

## Supplementary Table 11: Data Charting Form Extraction Fields and Rationale

| Data Element | Rationale for Extraction |
|---|---|
| Clinical context (setting, specialty, workflow, safety/governance) | Critical for understanding domain-specific challenges, generalizability, and necessary safety assessments in high-stakes clinical deployment |
| Task class (documentation, QA, diagnostic reasoning, etc.) | LaaJ effectiveness is proportional to task structure; necessary to synthesize where the judge achieves high agreement vs. where it struggles |
| Judge & generator identity (family/size; same vs. different) | Tracking judge models is necessary to assess the risk of self-preference bias and to analyze cost-vs-performance trade-offs |
| Evaluation format & rubric (pointwise/pairwise/ranking; criteria) | The paradigm dictates reliability; extraction confirms whether methods use pointwise scoring (prone to instability), pairwise comparison (prone to positional bias), or ranking methods |
| Adaptation (few-shot, SFT, DPO, RAG; chain-of-thought) | Domain-specific fine-tuning and CoT are crucial methodological innovations that enforce alignment with expert standards |
| Validation (human experts, credentials, agreement metrics) | Human comparison is the gold standard; extraction of metrics (ICC, κ, Pearson correlation) quantifies alignment between automated and expert judgment |

| Data Element | Rationale for Extraction |
|---|---|
| Reliability & bias (replicates, position/length/style/self-enhancement tests) | Testing for systemic biases is mandatory for a reliable clinical tool |
| Robustness (adversarial, rare diseases, atypical cases) | Measures the system's ability to maintain safe performance under challenging real-world or manipulated clinical inputs |
| Efficiency (latency, token cost vs. human time cost, failure rate) | Quantifies the scalability value proposition of LaaJ |
| Provenance (dataset size/source, de-identification, availability, languages) | Data source documentation is vital for reproducibility, ethical review, and assessing quality |
| Generalizability and deployment readiness | Tracks whether the judge system is feasible for deployment; ensemble design is tracked to analyze Multi-Judge Consensus effects |

## Supplementary Table 12: Application Domains and Clinical Task Distribution (N = 49)

| Category | Subcategory | n (%) | Description |
|---|---|---|---|
| **Clinical Content Evaluation** | | **17 (34.7%)** | |
| | Clinical documentation & summarization | 7 (14.3%) | Discharge summary evaluation, SOAP note generation, radiology report summarization |
| | Report generation & evaluation | 4 (8.2%) | Radiology report assessment, general AI output evaluation |
| | Information extraction & NER | 6 (12.2%) | Radiology info extraction, medical record anonymization, biomedical relation extraction |

| Category | Subcategory | n (%) | Description |
| --- | --- | --- | --- |
| **Medical Knowledge & Reasoning** | | **17 (34.7%)** | |
| | Medical QA & reasoning | 10 (20.4%) | General medical knowledge, specialty-specific assessment (ophthalmology, rheumatology, plastic surgery), statistical reasoning |
| | Diagnostic support | 6 (12.2%) | Safety benchmarks, OSCE-based assessment, diagnostic accuracy, psychiatric interviewing, thyroid malignancy classification |
| | Multi-task evaluation | 1 (2.0%) | Expert knowledge across dietetics and mental health |
| **Patient-Facing & Decision Support** | | **12 (24.5%)** | |
| | Patient education & communication | 6 (12.2%) | Patient information generation, text simplification, empathic communication, health coaching |
| | Clinical decision support | 6 (12.2%) | Medical consultation agents, contrast media risk assessment, dietary recommendations, RAG-based specialty QA |

| Category | Subcategory | n (%) | Description |
|---|---|---|---|
| **Other** | | **3 (6.1%)** | Acupuncture knowledge, general LaaJ validation framework, operational genetics knowledge base |

**Specialty distribution:** - Multiple specialties / general medicine: 22 (44.9%) - Radiology: 6 (12.2%) - Endocrinology / diabetology: 4 (8.2%) - Cardiology: 3 (6.1%) - Plastic / hand surgery: 3 (6.1%) - Psychiatry: 2 (4.1%) - Ophthalmology: 2 (4.1%)

*Notable gaps: emergency medicine, pediatrics, surgery beyond plastics.*

## Supplementary Table 13: MedJUDGE Framework Clinical Task Risk Stratification

| Risk Tier | Definition | Examples | Evidence from Review |
|---|---|---|---|
| **Tier A (Low)** | No direct patient impact; errors affect research or education only | Research QA, educational content grading, literature summarization | Most current studies (75.5%) operate here appropriately |
| **Tier B (Moderate)** | Indirect patient impact; mandatory human verification before any clinical action | Clinical documentation QA, EHR summary assessment, radiology report screening | Croxford et al. (ICC = 0.818), DOCLENS (ρ = 0.787) demonstrate feasibility |
| **Tier C (High)** | Direct patient care decisions; judge errors could cause patient harm | Triage support, diagnostic reasoning evaluation, medication safety alerts | Only GNQA deployed; most studies avoid this tier appropriately |

*These risk tiers map to established clinical governance structures: Tier A corresponds to IRB-exempt quality improvement; Tier B aligns with institutional clinical quality assurance; Tier C corresponds to clinical decision support tools potentially falling under FDA SaMD guidance.*

## Supplementary Table 14: MedJUDGE Framework Expert Validation Requirements by Risk Tier

| Risk Tier | Credential Requirement | Must Report |
|---|---|---|
| Tier A (Low) | Domain-relevant training | Expert–expert agreement baseline |
| Tier B (Moderate) | Board-certified or equivalent | Expert–expert AND judge–expert agreement; credential details |
| Tier C (High) | Subspecialty specialists | All above + cross-institution validation; separate development and validation panels |

## Supplementary Table 15: PRISMA-ScR Checklist

| Section | Item # | PRISMA-ScR Checklist Item | Reported | Location |
|---|---|---|---|---|
| **TITLE** | | | | |
| Title | 1 | Identify the report as a scoping review | Yes | Title: "A Scoping Review of LLM-as-a-Judge in Healthcare and the MedJUDGE Governance Framework" |
| **ABSTRACT** | | | | |
| Structured summary | 2 | Provide a structured summary including background, objectives, eligibility criteria, sources of evidence, charting methods, results, and conclusions | Yes | Abstract: databases (n=6), date range (Jan 2020–Jan 2026), screening (11,727 citations), inclusion (49 studies), key findings, and MedJUDGE framework |
| **INTRODUCTION** | | | | |

| Rationale | 3 | Describe the rationale for the review in the context of what is already known | Yes | Introduction ¶1–4: LaaJ adoption in healthcare, limitations of existing evaluation approaches, absence of governance-focused review |
|---|---|---|---|---|
| Objectives | 4 | Provide an explicit statement of the questions and objectives being addressed, with reference to the key elements of the PCC framework | Yes | Introduction (contributions 1–3); Methods (Table 7, PCC framework) |
| **METHODS** | | | | |
| Protocol and registration | 5 | Indicate whether a review protocol exists; provide registration number and registry name | N/A | No protocol was pre-registered. LLM-as-a-Judge in healthcare represents an emerging phenomenon with no prior scoping review; the review question, PCC framework, and eligibility criteria were developed iteratively as the landscape was characterized. Pre-registration is therefore not applicable. |
| Eligibility criteria | 6 | Specify characteristics of the sources of evidence used as eligibility criteria (e.g., years, language, publication status), and rationale | Yes | Methods (Study selection); Appendix B (Inclusion and Exclusion Criteria) |
| Information sources | 7 | Describe all information sources in the search (e.g., databases with date ranges, contact with authors) | Yes | Methods (Information sources): Ovid MEDLINE, EMBASE, Scopus, Web of Science, ACM Digital Library, IEEE Xplore; searched June 9, 2025 and January 28, 2026 |

| Section | # | Item | Reported | Location/Notes |
|---|---|---|---|---|
| Search | 8 | Present the full electronic search strategy for at least one database, including any limits used, such that it could be repeated | Yes | Appendix A: complete Boolean search strings for all six databases |
| Study selection | 9 | State the process for selecting sources of evidence (i.e., screening and eligibility) included in the scoping review | Yes | Methods (Study selection): two independent reviewers (TO, YR) for title/abstract; two reviewers (TO, CL) for full-text; third reviewer (CL) for arbitration |
| Data charting process | 10 | Describe the methods of charting data from the included sources of evidence (e.g., calibrated form, independently by two reviewers) | Yes | Methods (Data charting): two authors (CL, YW); Supplementary Table 11 (extraction fields) |
| Data items | 11 | List and define all variables for which data were sought and any assumptions made | Yes | Supplementary Table 11: 12 extraction domains including clinical context, judge/generator identity, adaptation methods, validation, bias testing, deployment readiness |
| Critical appraisal of individual sources of evidence | 12 | If performed, provide a rationale for conducting a critical appraisal of included sources of evidence; describe the methods used and how this information was used in any potential synthesis | Partial | Formal risk-of-bias scoring not applied, consistent with scoping review methodology. Validation rigor assessed descriptively across bias testing, human involvement, and deployment readiness dimensions (Results: Validation Rigors) |
| Synthesis of results | 13 | Describe the methods of handling and summarizing the data that were charted | Yes | Methods (Synthesis): three-phase narrative synthesis — descriptive statistics, thematic analysis, quality and gap assessment |
| **RESULTS** | | | | |

| Study selection | 14 | Give numbers of sources of evidence screened, assessed for eligibility, and included in the review, with reasons for exclusions at each stage — ideally using a flow diagram | Yes | Results (Study selection results): 12,379 citations identified; 652 duplicates removed; 11,727 screened; 33 from database search; 16 from citation tracking; 49 total included. Fig. 1 (PRISMA-ScR flow diagram) |
|---|---|---|---|---|
| Presentation of results | 15 | Present the characteristics of the included sources of evidence | Yes | Results: application categories, clinical tasks, evaluation paradigms, judge model selection, validation rigors; Supplementary Tables 1–14 |
| Summary of evidence | 16 | Summarize and/or present the charting results as they relate to the review objectives | Yes | Results (all subsections); Discussion (governance gaps, Hidden Judge Problem, Shared Ignorance Problem); MedJUDGE framework |
| Limitations | 17 | Describe the limitations of the scoping review process | Yes | Limitations: grey literature inclusion, 49-study corpus, January 2026 search cutoff, MedJUDGE as conceptual framework pending empirical calibration |
| Conclusions | 18 | Provide a general interpretation of the results with respect to the review objectives and potential implications for future research | Yes | Conclusion; Future Work (five priority directions) |
| **FUNDING** | | | | |
| Funding | 19 | Describe sources of funding or other support and the role of funders | Pending | To be added prior to submission |

## Supplementary Note 1: Heterogeneity of Clinical Reasoning

An additional dimension not captured by the structured-to-complex gradient reported in the main text is the heterogeneity of clinical reasoning itself. Clinical evaluation encompasses distinct cognitive modes that carry different error profiles and require different judge evaluation criteria:

- **Diagnostic reasoning** relies primarily on pattern recognition and differential generation
- **Prognostic reasoning** requires integration of temporal trajectories across multiple data streams
- **Therapeutic reasoning** involves risk-benefit balancing under uncertainty with patient-specific contextual factors
- **Care coordination reasoning** demands evaluation of sequencing, timing, and communication across providers and settings

An LLM judge that reliably evaluates whether a diagnosis is supported by available evidence may fail entirely when asked to evaluate whether the timing of a therapeutic intervention is appropriate or whether a care plan adequately coordinates across specialties. None of the 49 included studies decomposed clinical reasoning along these lines, and the field would benefit from validation studies that assess judge performance separately across these distinct reasoning modes, as each may require different evaluation rubrics, different expert validator compositions, and potentially different judge architectures.

## Supplementary Note 2: Self-Preference Amplification Loop in Category 2

The self-preference amplification loop is particularly acute in Category 2 (reinforcement learning) applications where judges serve as reward models. When the same model family generates clinical content and evaluates it, or when judges from similar training distributions are used, the following feedback cycle can emerge:

1. The judge systematically rates content from its own family more favorably
2. The favorable rating feeds into training as a reward signal
3. The generator learns to produce content that appeals to the judge's biases rather than content that is clinically optimal
4. The cycle repeats, amplifying evaluation errors rather than correcting them

This creates a closed feedback loop with no external anchor to clinical ground truth. Only 2 of 49 studies (4.1%) tested self-enhancement bias, leaving this risk unquantified for the majority of healthcare LaaJ applications.

## Supplementary Note 3: Industry Ecosystem and Commercial Platforms

LLM-as-a-Judge has emerged as a critical industry focus, with major technology platforms and specialized vendors developing evaluation frameworks:

| Platform | Description |
| --- | --- |
| Confident AI | LLM evaluation and testing platform with judge implementations |
| Hugging Face | Open-source LLM-as-a-judge cookbook and tools |
| LangChain / LangSmith | Evaluation framework with built-in judge templates |
| Evidently AI | LLM monitoring and evaluation platform |
| Arize Phoenix | LLM observability with judge-based evaluation |
| Microsoft Azure Databricks | Enterprise LLM judge capabilities |
| Humanloop | LLM evaluation platform for enterprises |
| Patronus AI | LLM testing and evaluation service |
| Langfuse | Open-source LLM evaluation with judge methods |
| Galileo AI | LLM evaluation and monitoring platform |

Healthcare-specific commercial deployments include John Snow Labs' prior authorization evaluation systems and numerous clinical documentation quality platforms. This industry ecosystem suggests practical LaaJ adoption substantially exceeds peer-reviewed literature. However, healthcare-specific implementations rarely disclose validation rigor, expert agreement metrics, bias testing results, or robustness assessments. Healthcare institutions considering commercial LaaJ services should demand vendor documentation meeting academic validation standards.

## Appendix A: Complete Search Strategy

### Ovid MEDLINE and EMBASE

Database(s): Embase 1974 to 2025 June 03, Ovid MEDLINE(R) and Epub Ahead of Print, In-Process, In-Data-Review & Other Non-Indexed Citations, Daily and Versions 1946 to June 03, 2025

| # | Searches | Results |
| --- | --- | --- |
| 1 | exp Large Language Models/ | 577 |
| 2 | (ChatGPT or "generative AI" or "generative artificial intelligence" or "generative language model" or "generative language models" or "generative pre-trained transformer" or GPT or GPT2 or "GPT-2" or GPT35 or "GPT-35" or GPT4 or "GPT-4" or "language model" or "language models" or "large language model" or "large language models" or LLM or LLMs).ti,ab,kf. | 45,253 |
| 3 | 1 or 2 | 45,254 |
| 4 | (eval or evaluate or evaluated or evaluates or evaluation or evaluations or evaluator or evaluators).ti,ab,kf. | 11,673,789 |
| 5 | 3 and 4 | 13,943 |
| 6 | limit 5 to english language | 12,871 |
| 7 | limit 5 to no language specified | 0 |
| 8 | 6 or 7 | 12,871 |
| 9 | limit 8 to yr="2020-Current" | 11,308 |
| 10 | ((meta adj analys*) or metaanalys* or (systematic* adj3 review*) or review or survey).ti,pt. | 8,648,720 |
| 11 | 9 not 10 | 10,421 |
| 12 | limit 11 to (letter or conference abstract or editorial or erratum or note…) | 1,585 |
| 13 | 11 not 12 | 8,836 |

| # | Searches | Results |
|---|---|---|
| 14 | limit 13 to yr="2025-Current" | 3,304 |
| 15 | remove duplicates from 14 | 1,962 |
| 16 | limit 13 to yr="2024" | 3,798 |
| 17 | remove duplicates from 16 | 2,387 |
| 18 | 13 not (14 or 16) | 1,734 |
| 19 | remove duplicates from 18 | 115 |

**Scopus**

1 TITLE(ChatGPT OR "generative AI" OR "generative artificial intelligence" OR "generative language model" OR "generative language models" OR "generative pre-trained transformer" OR GPT OR GPT2 OR "GPT-2" OR GPT35 OR "GPT-35" OR GPT4 OR "GPT-4" OR "language model" OR "language models" OR "large language model" OR "large language models" OR LLM OR LLMs)
2 TITLE(eval OR evaluate OR evaluated OR evaluates OR evaluation OR evaluations OR evaluator OR evaluators)
3 PUBYEAR AFT 2019 AND LANGUAGE(english)
4 1 and 2 and 3
5 TITLE((meta W/1 analys*) OR metaanalys* OR (systematic* W/3 review*) OR review)
6 4 and not 5
7 DOCTYPE(le) OR DOCTYPE(ab) OR DOCTYPE(ed) OR DOCTYPE(bk) OR DOCTYPE(er) OR DOCTYPE(no) OR DOCTYPE(sh)
8 6 and not 7
9 INDEX(embase) OR INDEX(medline) OR PMID(0* OR 1* OR 2* OR 3* OR 4* OR 5* OR 6* OR 7* OR 8* OR 9*)
10 8 and not 9

**Web of Science**

1 (ChatGPT OR "generative AI" OR ... OR LLM OR LLMs) (Title) and (eval OR evaluate OR ... OR evaluators) (Title) and English (Language) and Article OR Proceedings Paper (Document Type)
2 TI=(((meta NEAR/1 analys*) OR metaanalys* OR (systematic* NEAR/3 review*) OR review))
3 1 not 2
4 PMID=(0* or 1* or 2* or 3* or 4* or 5* or 6* or 7* or 8* or 9*)
5 3 not 4

**IEEE Xplore**

("Document Title":ChatGPT OR "Document Title":"generative AI" OR ... OR "Document Title":LLMs) AND ("Document Title":eval OR ... OR "Document Title":evaluators) NOT ("Document Title":"meta-analysis" OR "Document Title":metaanalysis OR "Document Title":"systematic review" OR "Document Title":review)
Limit 2020-2025

**ACM Digital Library**

[[Title: chatgpt] OR [Title: "generative ai"] OR ... OR [Title: llms]] AND [[Title: eval] OR ... OR [Title: evaluators]] AND [E-Publication Date: (01/01/2020 TO 06/30/2025)]

## Appendix B: Inclusion and Exclusion Criteria

**Inclusion Criteria**
1. The study uses Large Language Models (GPT, Claude, PaLM, LLaMA, or similar) as evaluators, judges, or assessors of content or performance
2. The evaluation/judgment task occurs in a medical or healthcare context (clinical practice, medical education, public health, healthcare administration, patient care, or medical research)
3. The publication presents original research, validation studies, proof-of-concept work, or case studies with actual implementation or testing
4. The publication is a peer-reviewed journal article, full conference paper, or preprint from recognized repositories

**Exclusion Criteria**
1. Studies where LLMs are only used for content generation without evaluation functions
2. Studies where LLMs are being evaluated rather than serving as evaluators
3. Studies without explicit healthcare or medical applications
4. Traditional NLP or machine learning studies that do not involve LLMs
5. Editorials, commentaries, opinion pieces, conference abstracts only, dissertations, books, or non-academic publications
6. Non-English publications
7. Duplicate publications of the same study

**Screening Instructions**
- Include if both LLM evaluation/judgment AND healthcare context are clearly mentioned
- Include for full-text review if uncertain based on title/abstract
- Include studies where LLMs are part of hybrid human-AI evaluation systems in healthcare
- Include studies comparing LLM judgments to human healthcare expert judgments

## Appendix C: Progressive Keyword Screening Results (N = 11,727 after deduplication)

| Step | Keyword Layer | Pattern | New | Cumulative | + Healthcare |
|---|---|---|---|---|---|
| 1 | Exact "LLM as a judge" | llm as a judge | 54 | 54 | 17 |
| 2 | + LLM/langua | LLM judge, language | 11 | 65 | 19 |

| Step | Keyword Layer | Pattern | New | Cumulative | + Healthcare |
|---|---|---|---|---|---|
| | ge model judge | model judge, AI judge | | | |
| 3 | + LLM/GPT as evaluator/assessor | LLM as evaluator, GPT as scorer/rater/annotator | 2 | 67 | 20 |
| 4 | + LLM/GPT-based evaluation | LLM-based evaluation, GPT-based scoring | 36 | 103 | 36 |
| 5 | + Automated eval using LLM | automated evaluation using/with LLM/GPT | 15 | 118 | 42 |
| 6 | + Evaluation framework + LLM | evaluation framework LLM, benchmark framework | 3 | 121 | 44 |
| 7 | + Quality/text eval + automated | quality evaluation automated, text scoring LLM | 139 | 260 | 139 |
| 8 | + G-Eval | G-Eval (NLG evaluation method) | 218 | 478 | 236 |
| 9 | + RLHF / reward model / DPO | reward model, RLHF, RLAIF, preference optimization | 18 | 496 | 243 |
| 10 | + Pointwise/pairwise evaluation | pointwise evaluation, pairwise | 31 | 527 | 268 |

| Step | Keyword Layer | Pattern | New | Cumulative | + Healthcare |
|---|---|---|---|---|---|
| | | comparison/ranking | | | |
| 11 | + Multi-agent/agent-based eval | multi-agent evaluation, agent-based assessment | 61 | 588 | 298 |
| 12 | + Inter-rater agreement + LLM | inter-rater agreement LLM, annotator agreement AI | 455 | 1,043 | 720 |
| 13 | + LLM-expert agreement | LLM-human agreement, model-clinician concordance | 1 | 1,044 | 721 |
| **TOTAL** | | | | **1,044** | **721** |

# Supplementary Figures

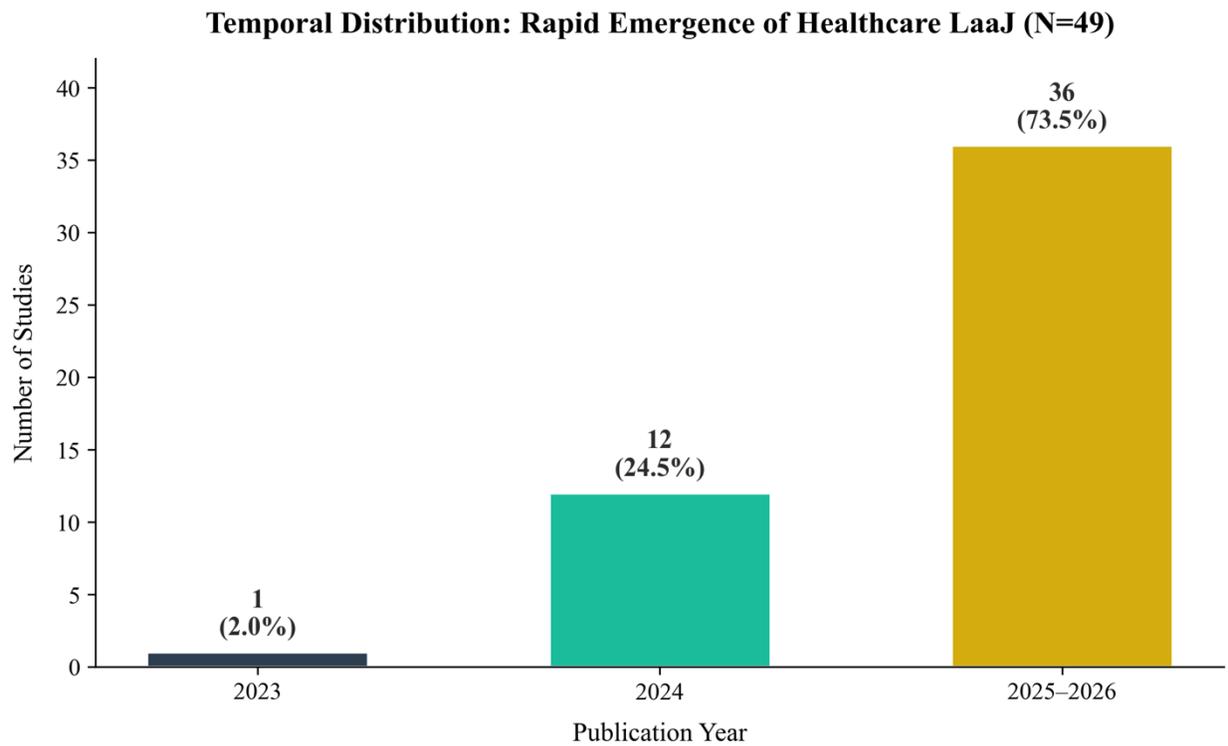

**Supplementary Fig. 1 | Temporal distribution of included studies (N=49).** Bar chart showing the publication year distribution of healthcare LaaJ studies, demonstrating the rapid emergence from 2023 to 2026. The exponential growth reflects the field's nascent but accelerating adoption. Source file:

figures/supplementary/Supplementary_Fig1_Year_Distribution.png.

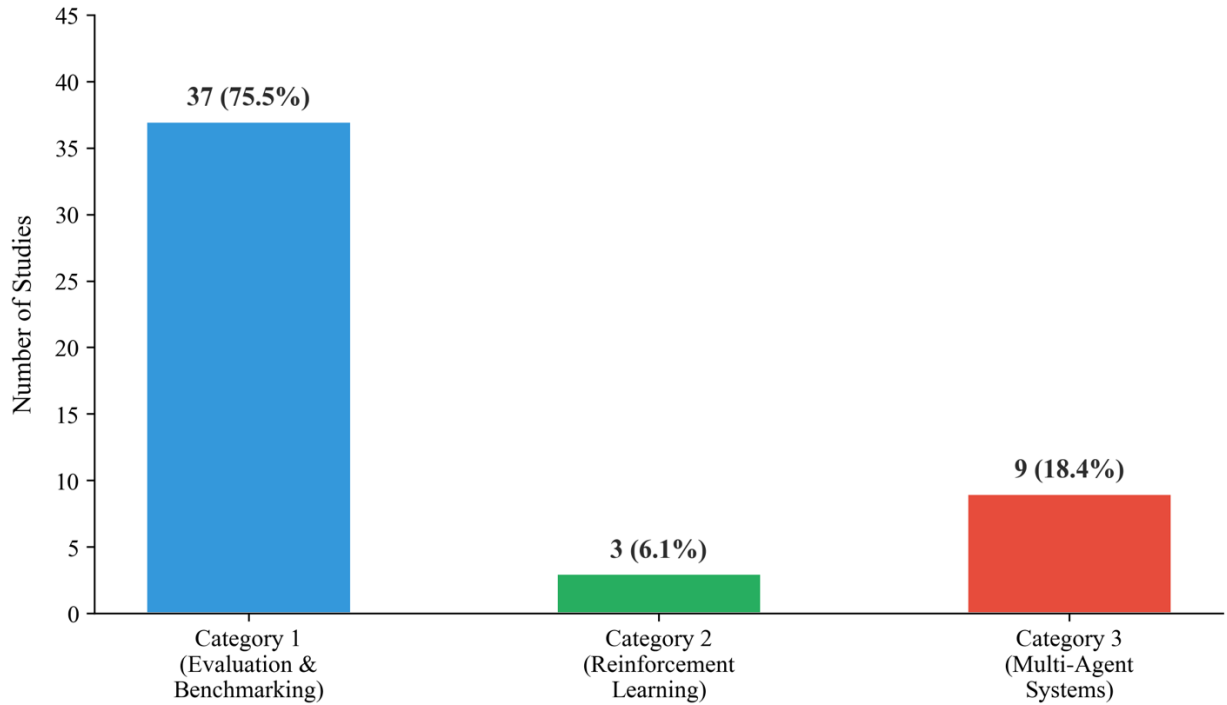

**Supplementary Fig. 2 | Distribution of studies across application tiers (N=49).** Bar chart showing the distribution of included studies across three application categories: Tier 1 (Evaluation & Benchmarking, 37 studies, 75.5%), Tier 2 (Reinforcement Learning, 3 studies, 6.1%), and Tier 3 (Multi-Agent Systems, 9 studies, 18.4%). Source file: figures/supplementary/Supplementary_Fig2_Category_Distribution.png.

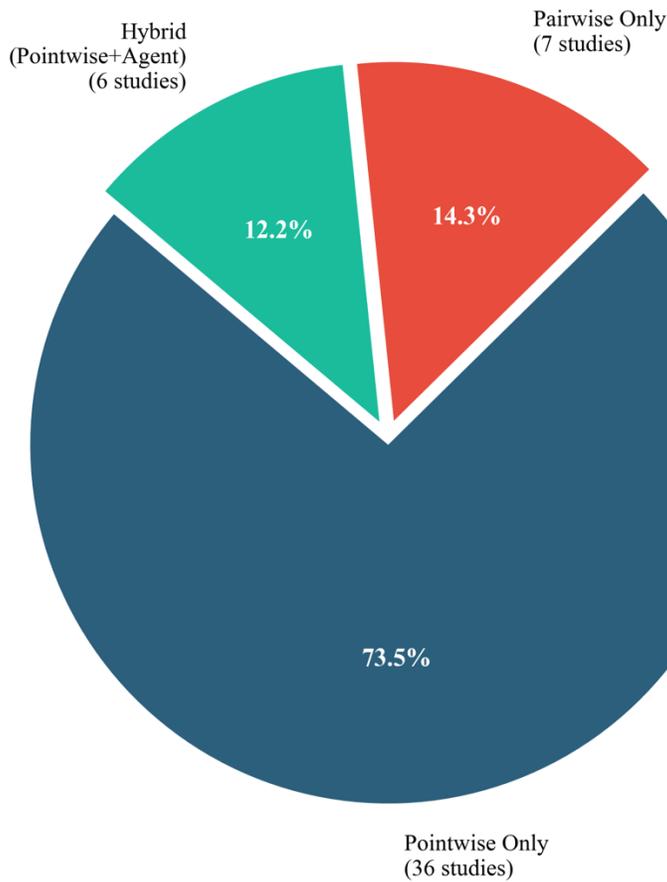

**Supplementary Fig. 3 | Evaluation paradigm distribution (N=49).** Pie chart illustrating the distribution of evaluation paradigms across included studies. Pointwise scoring was used in 42 studies (85.7%), pairwise comparison in 7 (14.3%), and agent-based evaluation in 6 (12.2%); categories overlap as 6 hybrid studies employed both pointwise and agent-based approaches. The chart shows mutually exclusive slices: pointwise only (73.5%), pairwise only (14.3%), and hybrid pointwise+agent (12.2%). Source file: figures/supplementary/Supplementary_Fig3_Paradigm_Distribution.png.

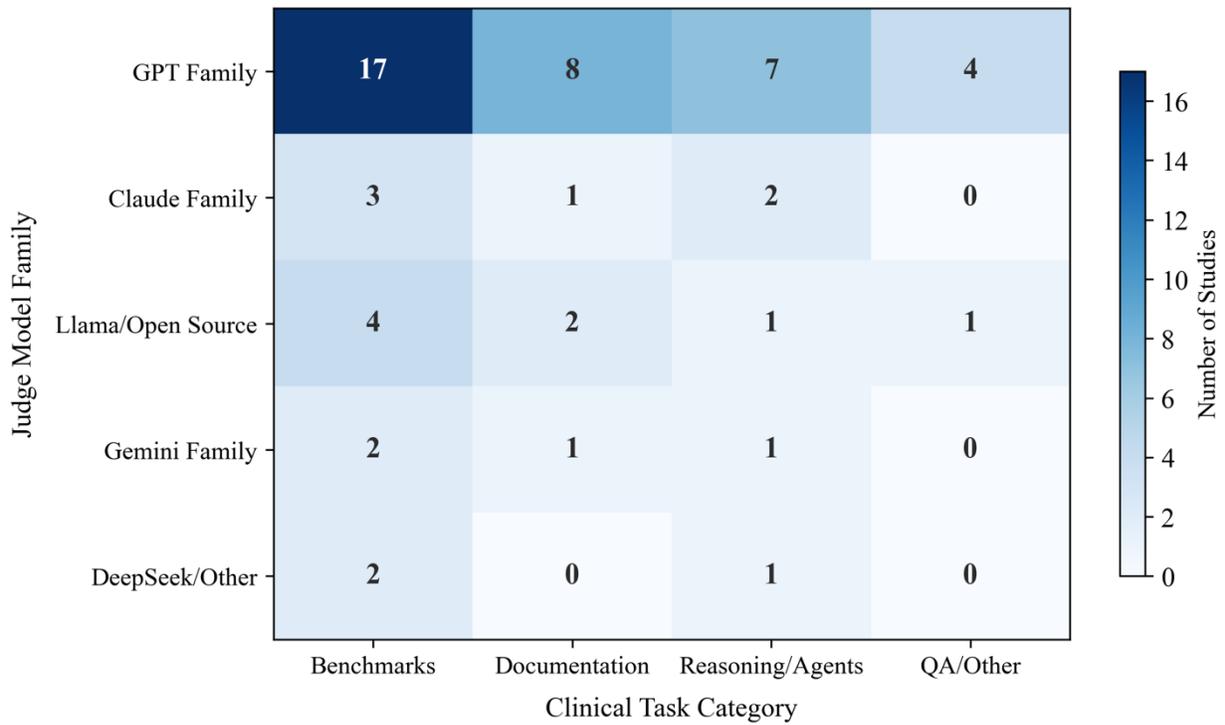

**Supplementary Fig. 4 | Distribution of judge model architectures across clinical tasks (N=49).** Heatmap showing the relationship between judge model families (GPT, Claude, Llama/Open Source, Gemini, DeepSeek/Other) and clinical task categories (Benchmarks, Documentation, Reasoning/Agents, Q&A/Other). GPT-family models dominate across all task categories, with the highest concentration in benchmarking tasks (18 studies). Source file: figures/supplementary/Supplementary_Fig4_Judge_Landscape_Heatmap.png.

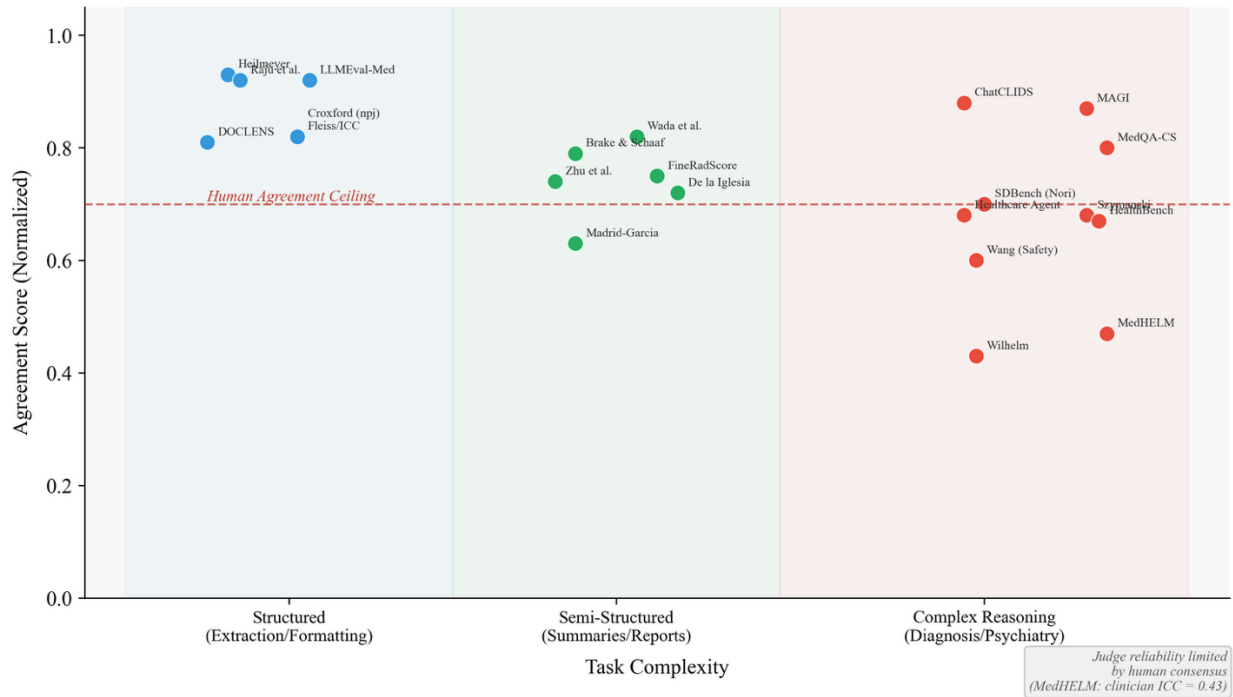

**Supplementary Fig. 5 | Bias testing audit card: critical gaps in healthcare LaaJ validation (N=49).** Horizontal bar chart showing the percentage of studies conducting each type of bias testing, with associated clinical risk levels. Positional bias (8.2%), cross-specialty bias (6.1%), and self-enhancement (4.1%) were the only dimensions tested by more than one study. Critical-risk dimensions—demographic (race, gender, SES), severity calibration, and temporal stability,had 0% testing rates. Source file: figures/supplementary/Supplementary_Fig8_Bias_Testing_Audit.png.

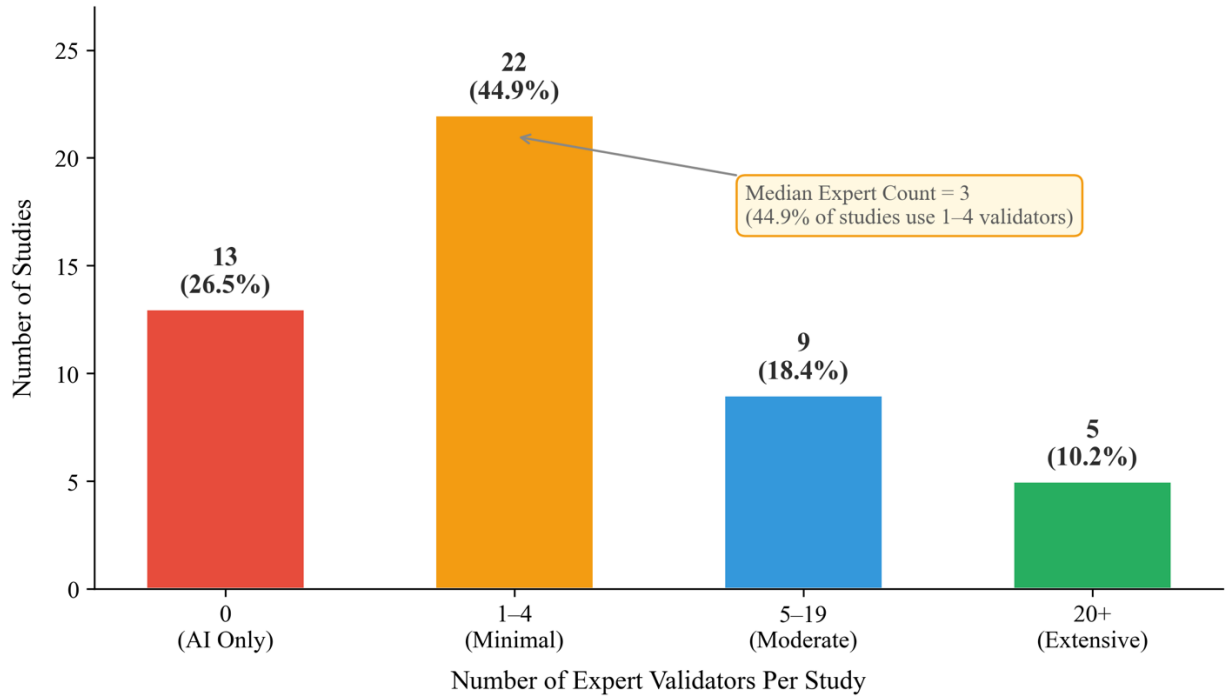

**Supplementary Fig. 6 | The validation crisis: distribution of expert validators per study (N=49).** Histogram showing the number of expert validators used across included studies: 0 experts/AI only (13 studies, 26.5%), 1–4 experts (21 studies, 42.9%), 5–19 experts (10 studies, 20.4%), and 20+ studies, 10.2%). The median expert count is 3, highlighting the limited human oversight in current LaaJ validation practices. Source file: figures/supplementary/Supplementary_Fig6_Validation_Crisis.png.

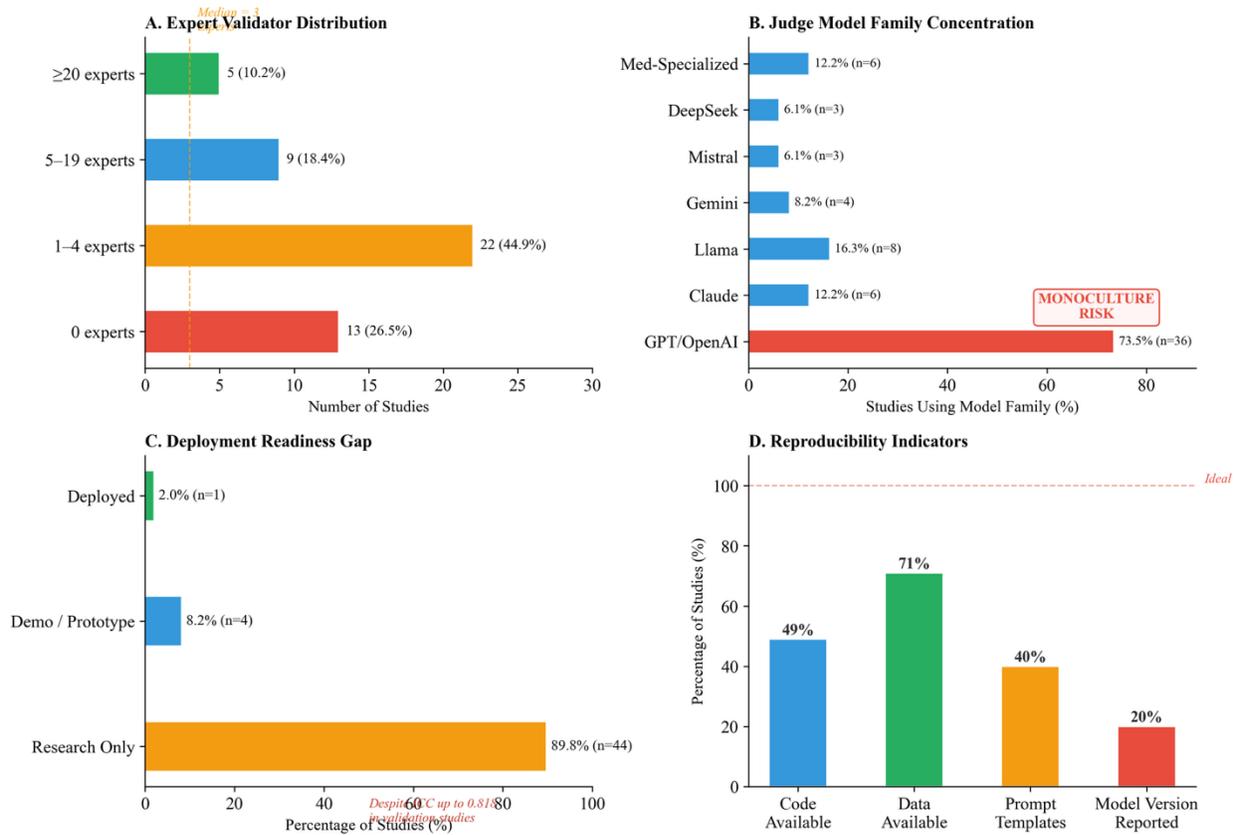

**Supplementary Fig. 7 | Reliability versus complexity tradeoff in healthcare LaaJ (N=22).** Scatter plot of agreement scores (normalized) versus task complexity (Structured, Semi-Structured, Complex Reasoning) for studies reporting quantitative validation metrics. Studies are colored by task category. The human agreement ceiling line (approximately 0.7) indicates the upper bound of inter-rater reliability. Complex reasoning tasks show the widest spread in agreement scores, with some systems falling below the human agreement ceiling. Source file: figures/supplementary/Supplementary_Fig5_Reliability_Complexity_Tradeoff.png.

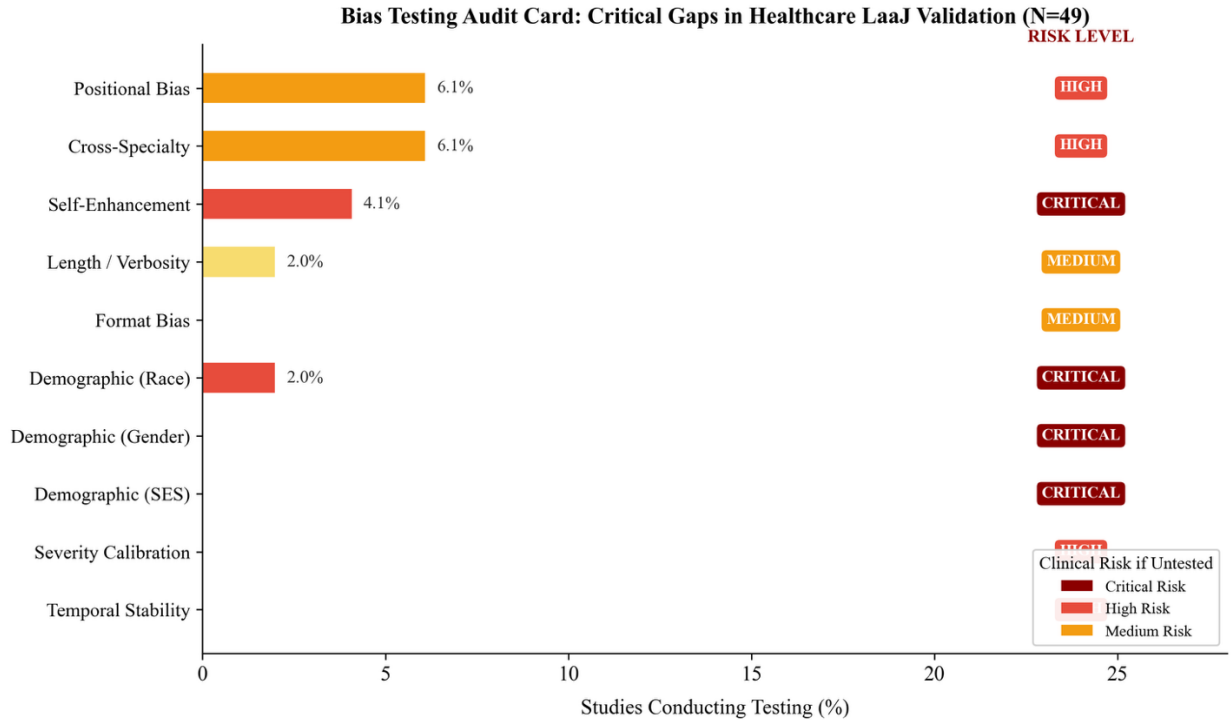

**Supplementary Fig. 8 | Validation rigor gaps in healthcare LaaJ (N=49).** Four-panel composite figure showing: (A) Expert validator distribution matching Supplementary Table 9 (0 experts: 13 studies, 1–4: 21, 5–19: 10, ≥20: 5), (B) Judge model family concentration revealing GPT/OpenAI monoculture risk (75.5%, n=37), with Claude (14.3%, n=7), Llama (16.3%, n=8), and Gemini (10.2%, n=5), (C) Deployment readiness gap between research-only (89.8%, n=44), demo/prototype (8.2%, n=4), and deployed systems (2.0%, n=1), and (D) Reproducibility indicators showing low rates of code availability (49%), data availability (71%), prompt template sharing (40%), and model version reporting (20%). Source file: figures/supplementary/Supplementary_Fig7_Validation_Rigor_Gaps.png.

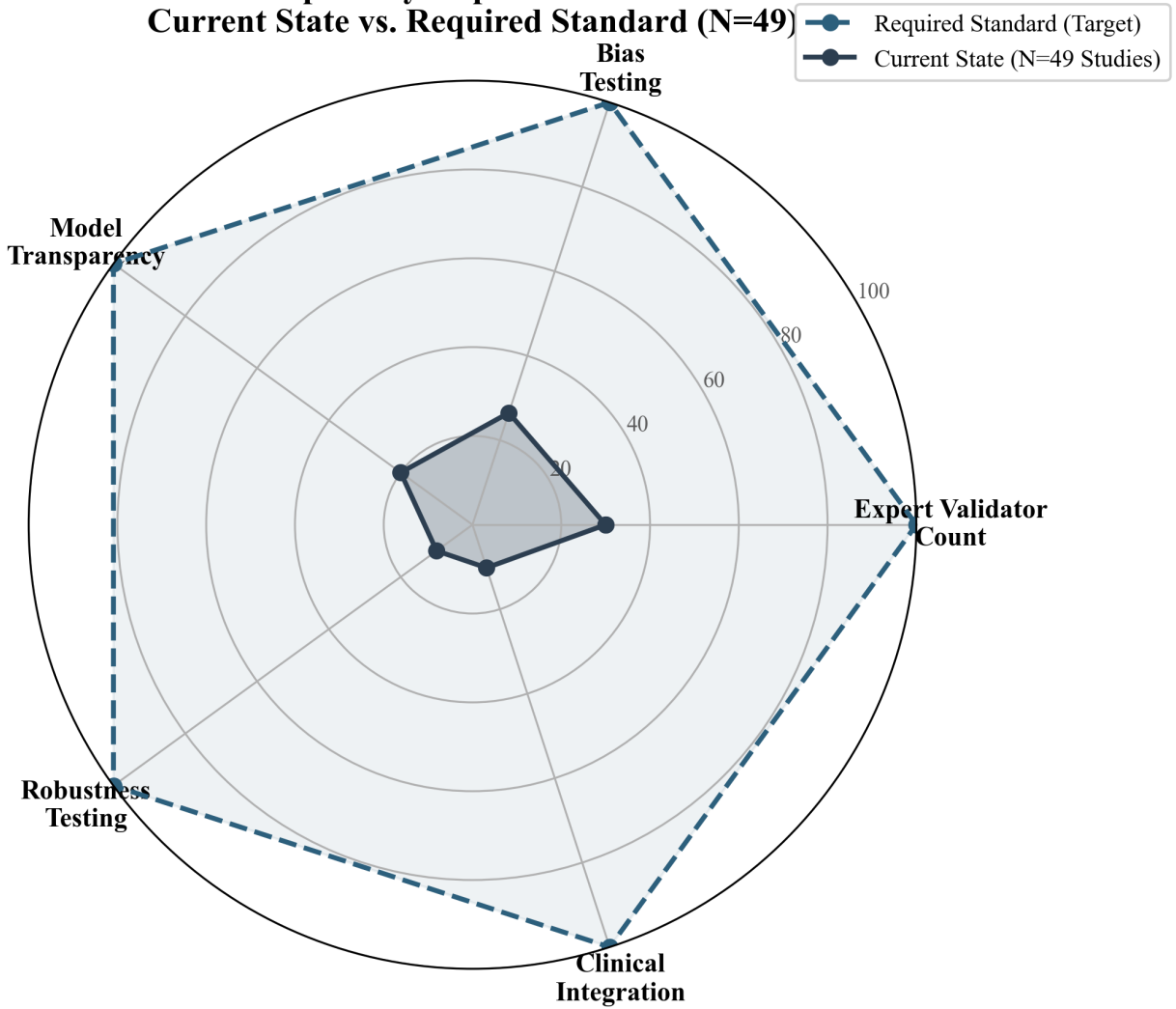

**Supplementary Fig. 9 | Transparency gap in healthcare LaaJ (N=49).** Radar chart comparing the current state of reporting practices against required standards across six dimensions: Expert Validator Count, Bias Testing, Model Transparency, Robustness Testing, and Clinical Integration. The large gap between the current state (inner polygon) and the required standard (outer dashed line) illustrates the systemic transparency deficit in current healthcare LaaJ research. Source file: figures/supplementary/Supplementary_Fig9_Transparency_Gap_Radar.png.